%% file: ms.tex
\documentclass[12pt,preprint]{aastex}

\newcommand{\pks}{PKS~2155--304}
\newcommand{\rmax}{$r_{\rm max}$}
\newcommand{\tpeak}{$\tau_{\rm peak}$}
\newcommand{\tcent}{$\tau_{\rm cent}$}
\newcommand{\tfit}{$\tau_{\rm fit}$}
\newcommand{\tcool}{$t_{\rm cool}$}
\newcommand{\tacc}{$t_{\rm acc}$}

\newcommand{\fvar}{$F_{\rm var}$}

\newcommand{\cha}{{\it Chandra}}
\newcommand{\xmm}{{\it XMM-Newton}}
\newcommand{\sax}{{\it BeppoSAX}}
\newcommand{\asca}{{\it ASCA}}
\newcommand{\xte}{{\it RXTE}}

\newcommand{\et}{et al.\ }

\slugcomment{accepted by ApJ} 

\shortauthors{Zhang et al. }
\shorttitle{XMM-Newton View of PKS~2155--304}

\begin{document}

\title{XMM-Newton View of PKS~2155--304: Hardness Ratio and Cross-correlation Analysis of EPIC pn Observations}

\authoraddr{ }

\author{Y.H. Zhang\altaffilmark{1},
        A. Treves\altaffilmark{2},
        L. Maraschi\altaffilmark{3},
        J.M. Bai\altaffilmark{4}, and 
	F.K. Liu\altaffilmark{5}
}

\altaffiltext{1}{Department of Physics and Tsinghua Center for Astrophysics (THCA), Tsinghua University, Beijing 100084, China; youhong.zhang@mail.tsinghua.edu.cn}
\altaffiltext{2}{Dipartimento di Scienze, Universit\`a degli Studi dell'Insubria, via Valleggio 11, I-22100 Como, Italy}
\altaffiltext{3}{Osservatorio Astronomico di Brera, via Brera 28, I-20121 Milano, Italy}
\altaffiltext{4}{National Astronomical Observatories/Yunnan Observatory, Chinese Academy of Sciences, P. O. Box 110, Kunming, Yunnan, 650011, China}
\altaffiltext{5}{Department of Astronomy, Peking University, Beijing 100871, China}

\begin{abstract}

With currently available \xmm\ EPIC pn observations spanned over about 3 years, we present a detailed spectral and temporal variability of the 0.2--10~keV X-ray emission from the X-ray bright BL Lac object \pks. The spectral variability is examined with a model independent hardness ratio method. We find that the spectral evolution of the source follows the light curves well, indicating that the spectra harden when the fluxes increase. The plots of hardness ratios versus count rates show that the spectral changes are particularly significant during flares. The cross-correlation functions (CCFs) show that the light curves in the different energy bands are well correlated at different time lags. The CCF peaks (i.e., the maximum correlation coefficients) tend to become smaller with larger energy differences, and the variability in the different energy bands are more correlated for the flares than for the other cases. In most cases the higher energy band variations lead the lower energy band, but in two cases we observed the opposite behavior that the lower energy variability possibly leads the higher energy variability. The time lags increase with the energy differences between the two cross-correlated light curves. The maximum lag is found to be up to about one hour, in support with the findings obtained with previous low Earth orbit X-ray missions. We discuss our results in the context of the particle acceleration, cooling and light crossing timescales.

\end{abstract}



\section{Introduction}\label{sec:intro}

Blazars, including BL Lac objects and flat-spectrum radio quasars (FSRQs), are thought to be extragalactic sources with relativistic jets roughly aligned with the line of sight (e.g., Urry \& Padovani 1995). The multi-wavelength observations suggest that the electromagnetic emissions from blazars are dominated by non-thermal radiation produced by relativistic electrons tangled with magnetic field in the jets. Rapid and large amplitude variability is a remarkable characteristic of blazars, which have been observed on different timescales across the whole electromagnetic spectrum (for a review see Ulrich, Maraschi \& Urry 1997). Six blazars, usually designated as TeV blazars, have been detected up to TeV energies (see Krawczynski \et [2004] for a summary of the properties of the established TeV blazars including \pks). The TeV sources have been followed with special interests in both the observational and theoretical aspects since last decade. The spectral energy distributions (SEDs) of TeV blazars are well described by two overall emission components: synchrotron component peaking at UV/soft X-ray band, and inverse Compton component peaking at TeV energies. Simultaneous multi-wavelength observations showed that the variability around the two peaks are well correlated (e.g., Maraschi \et 1999 for Mrk~421), while recent observations of \pks\ did not find any correlations between the small ranges of the flux variations at different wavelengths (Aharonian \et 2005).

X-ray emissions from TeV blazars are known to be violently variable because they are thought to be the high energy tail of the synchrotron emission produced at the base of the relativistic jets. TeV blazars are therefore X-ray sources dominated by jet/synchrotron emission. Well-known TeV blazars are the three prototypical BL Lac objects Mrk~421, Mrk~501, and PKS~2155$-$304, which are important monitoring targets of almost all the X-ray missions. A number of observations have revealed very complex patterns of the temporal and spectral variability (see Pian 2002 and Zhang 2003 for reviews). The X-ray flares generally occurred on daily timescales as viewed from the long-look observations with \asca, \sax\ and \xte\ (e.g., Tanihata \et 2001; Zhang \et 2002; Massaro \et 2004a), while small amplitude and rapid flares were also frequently detected on timescales down to less than an hour (e.g., Cui 2004 for Mrk~421; Catanese \& Sambruna 2000, Xue \& Cui 2005 for Mrk~501; Aharonian \et 2005 for \pks). The X-ray variability of TeV blazars appears to be aperiodic and unpredictable in, e.g, flare intensity, duration and occurrence. Characterizing the X-ray variability of TeV blazars is therefore important for exploring the underlying physical process at work and the particle acceleration and cooling mechanisms.

The X-ray variations in the different energy bands are generally well correlated at different time lags (For Mrk~421 see e.g., Ravasio \et 2004; For Mrk~501 see e.g., Tanihata \et 2001; For \pks\ see e.g., Zhang \et 2002). The amounts of lags generally differ from flare to flare, and appear to be correlated with the flare durations: the larger the flare duration, the larger the lag (Zhang \et 2002; Brinkmann \et 2003). The delays can be either too small to be detectable or large up to a couple of hours. In most cases the lower energy variations lag behind the higher energy variations, while in some cases the lower energy variations lead ahead the higher energy variations. The delays increase with increasing differences of the compared energy bands (e.g., Zhang \et 2002). Preliminary evidence showed that in Mrk~421 the lags become larger with decreasing Fourier frequencies or increasing timescales (Zhang 2002). Due to limitations of observation lengths, the power spectral density (PSD) of TeV blazars were derived only in the high frequency range roughly between $10^{-5}$ and $10^{-3}$~Hz, and most likely described by a power-law ($\mathcal{P}(f) \propto f^{-\alpha}$, where $\mathcal{P}(f)$ is the power at frequency $f$) with slope $\alpha \sim 2-3$ (e.g, Zhang \et 1999, 2002; Zhang 2002; Brinkmann \et 2003), indicating that the variability amplitude rapidly decreases toward short timescales. In fact, the PSD shapes are not well known at low frequency while a break to $\alpha \sim 1$ at timescales of the order of days was suggested (Kataoka \et 2001). The X-ray spectra are better represented by complicated models (e.g., logarithmic parabola) rather than by a single power law or broken power law models, indicating that the X-ray spectra continuously steepen with increasing energies (Fossati \et 2000;  Massaro \et 2004a). The X-ray spectral index and the SED peak energy generally correlate with the source intensity (For Mrk~421 see Fossati \et 2000, Tanihata \et 2004;  For Mrk~501 see Massaro \et 2004b; For \pks\ see Zhang \et 2002).       

\pks\ is one of the best studied blazars in the X-rays. It has been repeatedly observed by previous X-ray missions, and now is one of the calibration sources for \cha\ and \xmm. A detailed analysis with three long \sax\ observations showed that the X-ray spectra of \pks\ are generally steeper, and hence the peak energies are smaller, than those of Mrk~421 and Mrk~501. As a calibration source, \pks\ has been observed by \xmm\ during 6 orbits spanned over about 3.5 years. Using the excess variance method, Zhang \et (2005, Paper I hereafter) have presented a detailed analysis of the X-ray variability characteristics. They found that the source variability is not a stationary process, mainly characterized by the fact that the fractional $rms$ variability amplitudes (\fvar) between 50 and 500~s become smaller when the source fluxes increase. The $rms$ spectra showed that the fractional $rms$ variability amplitudes exponentially increase with energies. In this paper we will use the same filtered event list datasets as presented in Paper I to further study the temporal and spectral variability of \pks. For the details of the observational journal and data reduction we refer the reader to Paper I. 

In \S~\ref{sec:hr} we show the normalized light curves and perform a model independent hardness ratio analysis for studying spectral variability of the source. Cross-correlation function analysis is described in \S~\ref{sec:ccf} to examine the correlations and to find possible time lags between the variability in the different energy bands. We discuss the physical implications of our results in \S~\ref{sec:disc}. The conclusions are given in \S~\ref{sec:conc}.

         
\section{Hardness Ratio Analysis}\label{sec:hr}

\begin{figure}
\plottwo{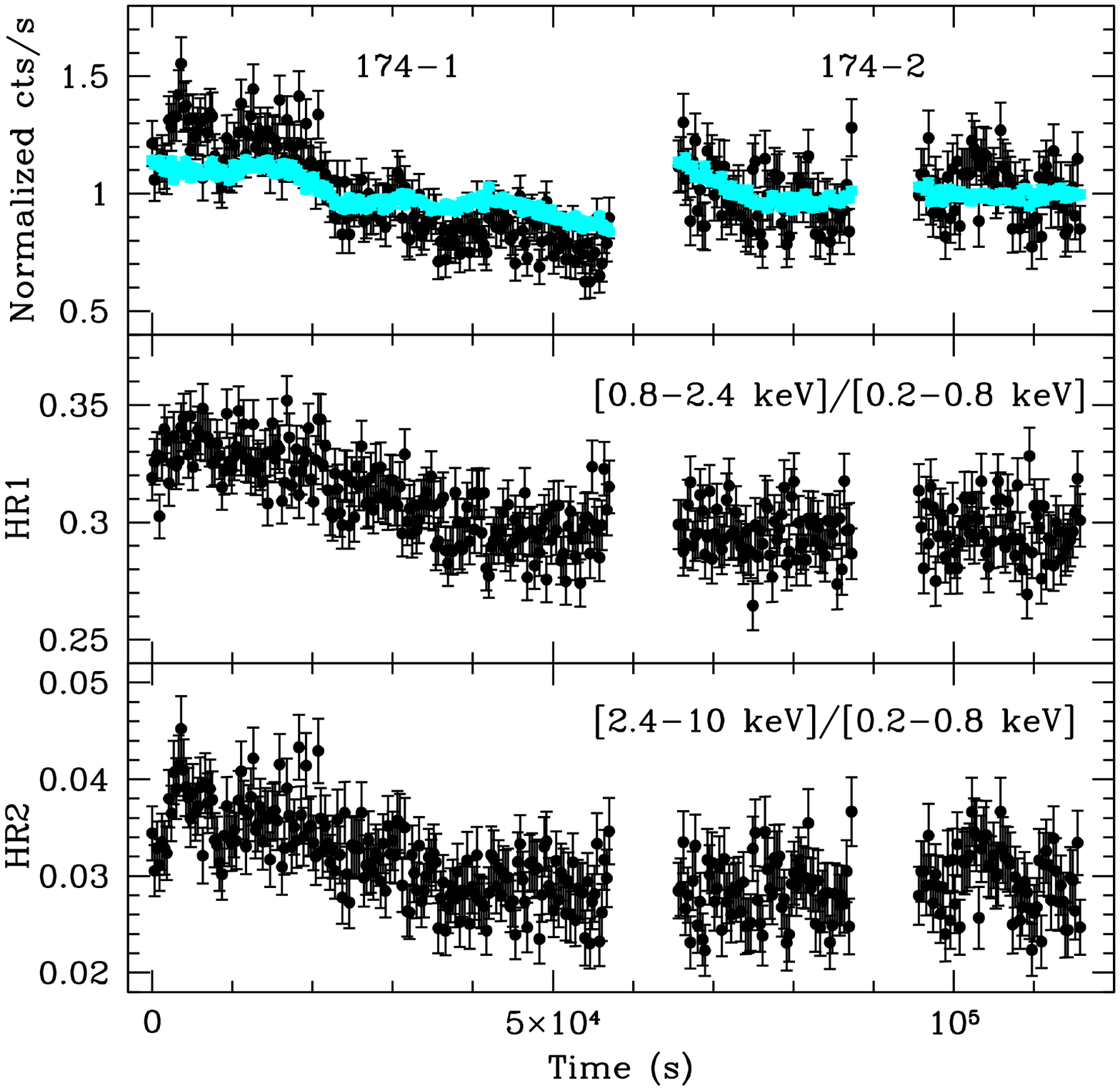}{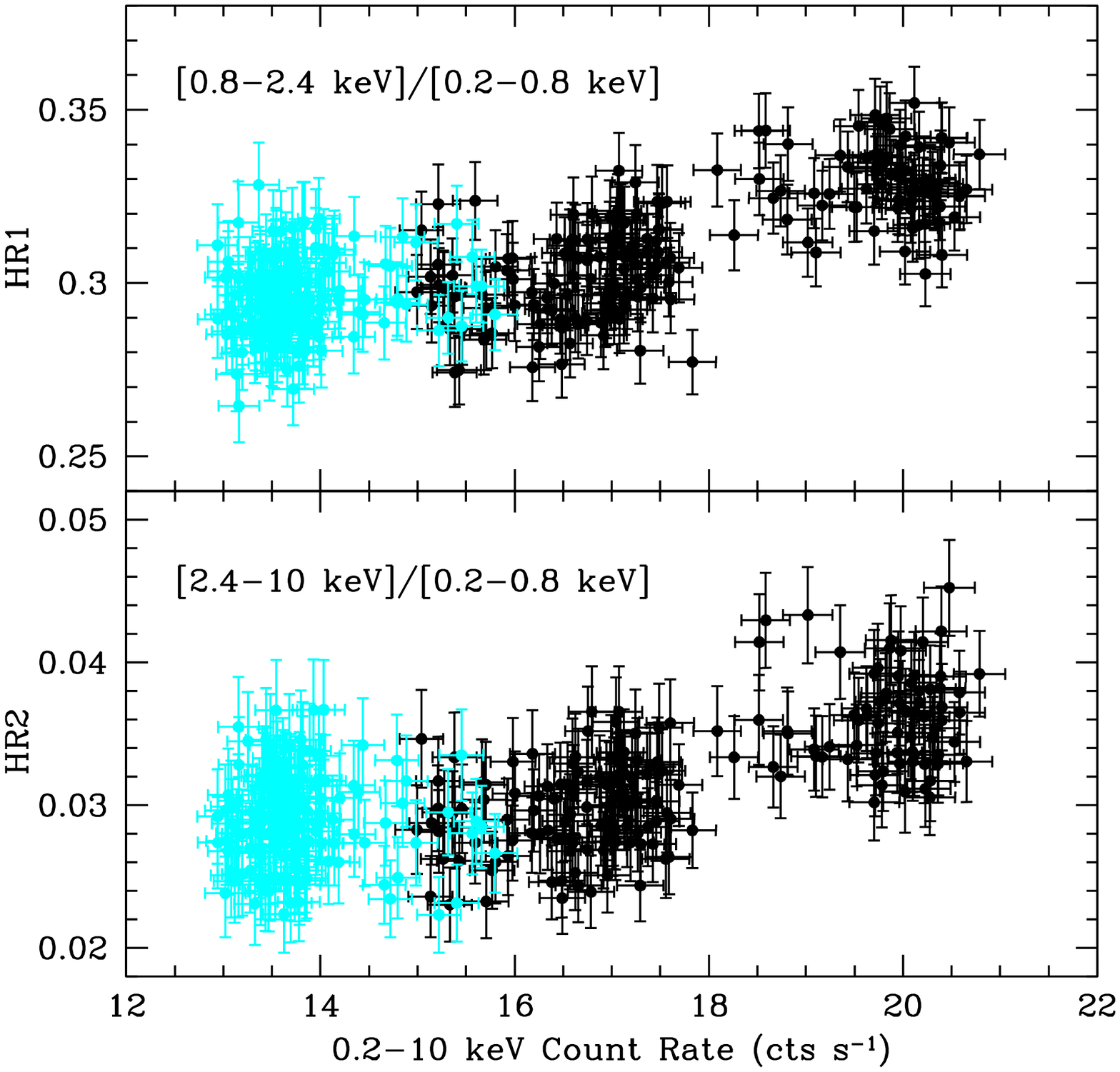}
\caption { \footnotesize 
For the Orbit 174 observation. 
Left picture: the upper panel plots the normalized 0.2--0.8~keV (grey, cyan in color) and the 2.4--10~keV (dark) energy band light curves. The offsets of the normalized light curves between the exposure 174-1 and 174-2 are caused by the different normalization with their own mean count rates, respectively. The medium and bottom panels plot the 0.8--2.4~keV/0.2--0.8~keV hardness ratios (HR1) and the 2.4--10~keV/0.2--0.8~keV hardness ratios (HR2) as a function of time, respectively. Both the light curves and hardness ratios are binned in 300~s. 
Right picture: the upper and bottom panels plot HR1 and HR2 as a function of the total 0.2--10~keV count rates, respectively (the exposure 174-1 is in dark and the exposure 174-2 in grey). Note that the different extraction regions yield small offset of the 0.2--10~keV count rates between the two exposures. }
\label{fig:lc:174}
\end{figure}

\begin{figure}
\plottwo{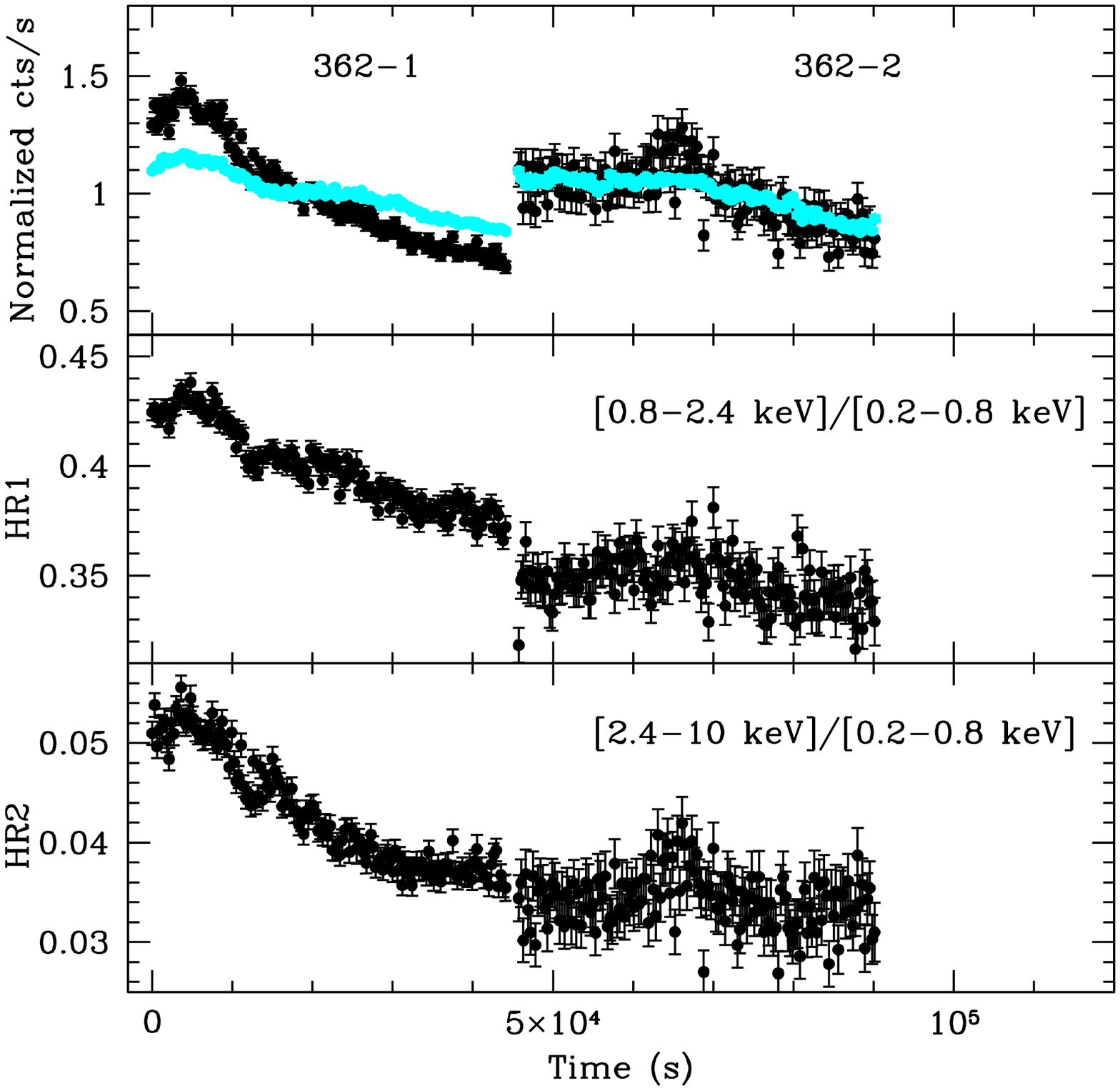}{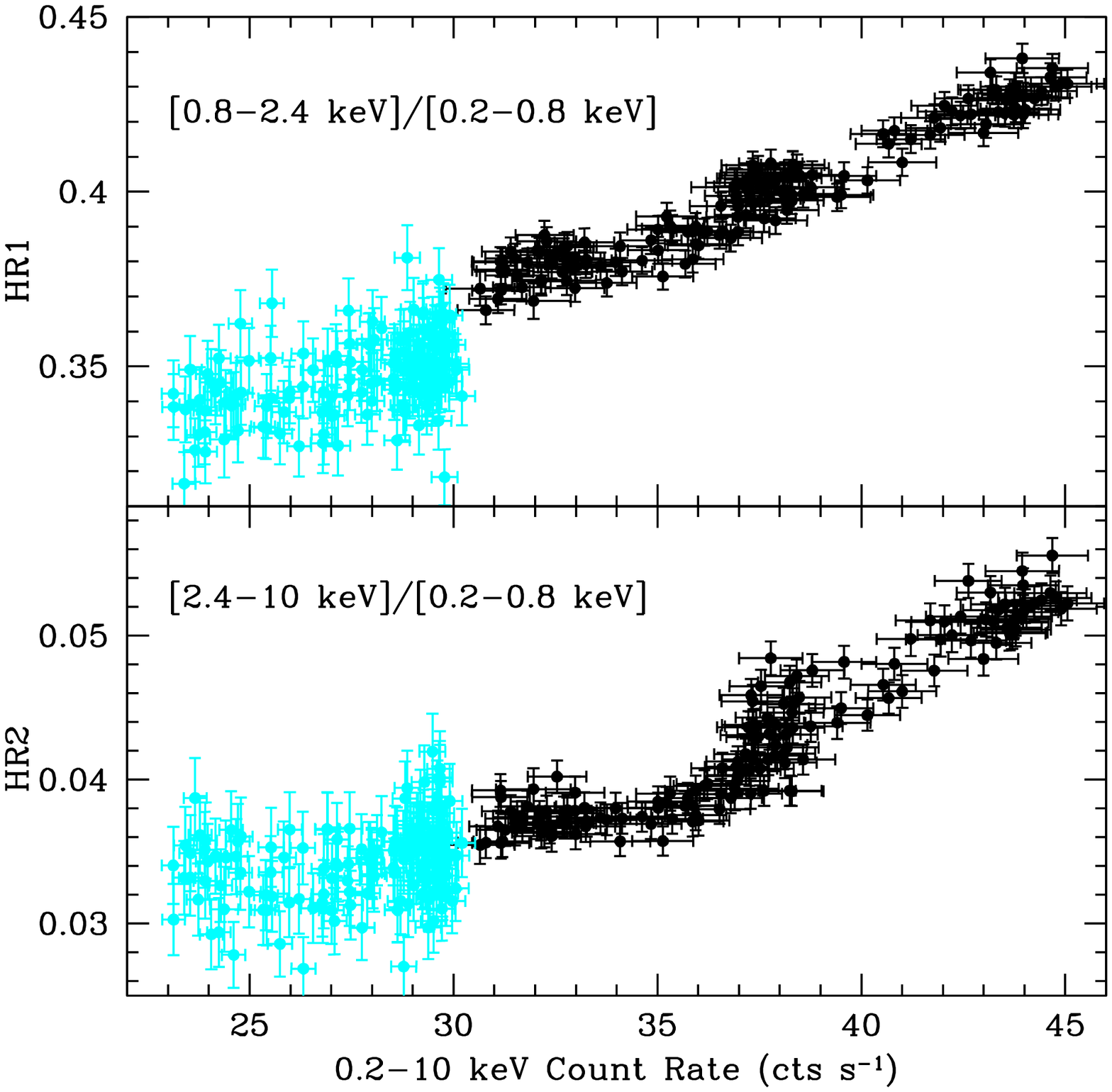}
\caption { \footnotesize As Figure~\ref{fig:lc:174} but for the Orbit 362 observation. In the left picture, the offsets of the normalized light curves (upper panel) between the exposure 362-1 (in timing mode) and 362-2 (in imaging mode) are resulted from the different normalization with their own mean count rates, respectively. The different operating modes also yield the offsets in the hardness ratios (particularly significant in HR1) between the two exposures as seen from the medium and bottom panels. In the right picture, the count rates of the exposure 362-1 (dark) are arbitrarily scaled down by a factor of 4.75 for smoothly connecting with those of the exposure 362-2 (grey). } 
\label{fig:lc:362}
\end{figure}

\begin{figure}
\plottwo{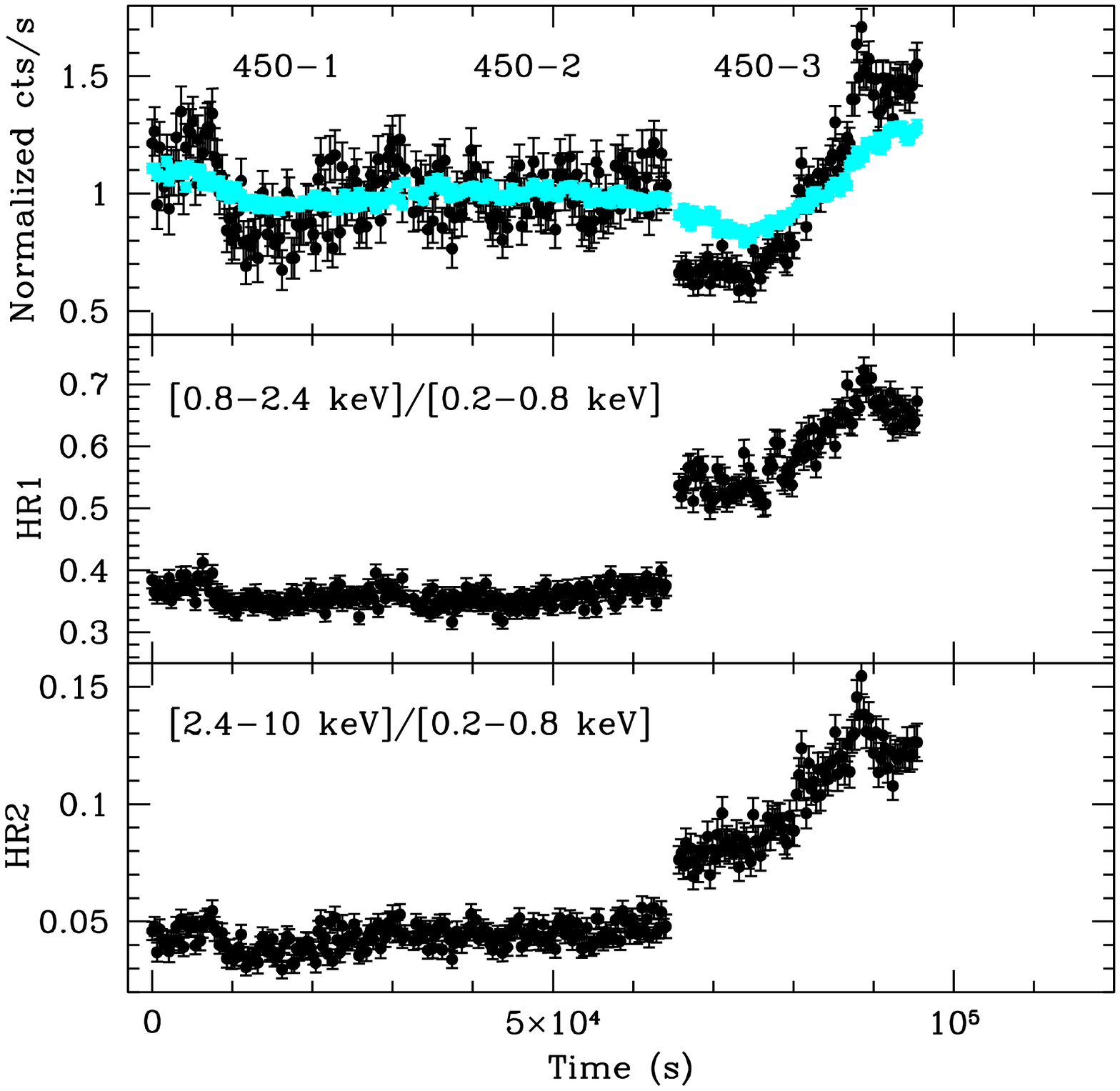}{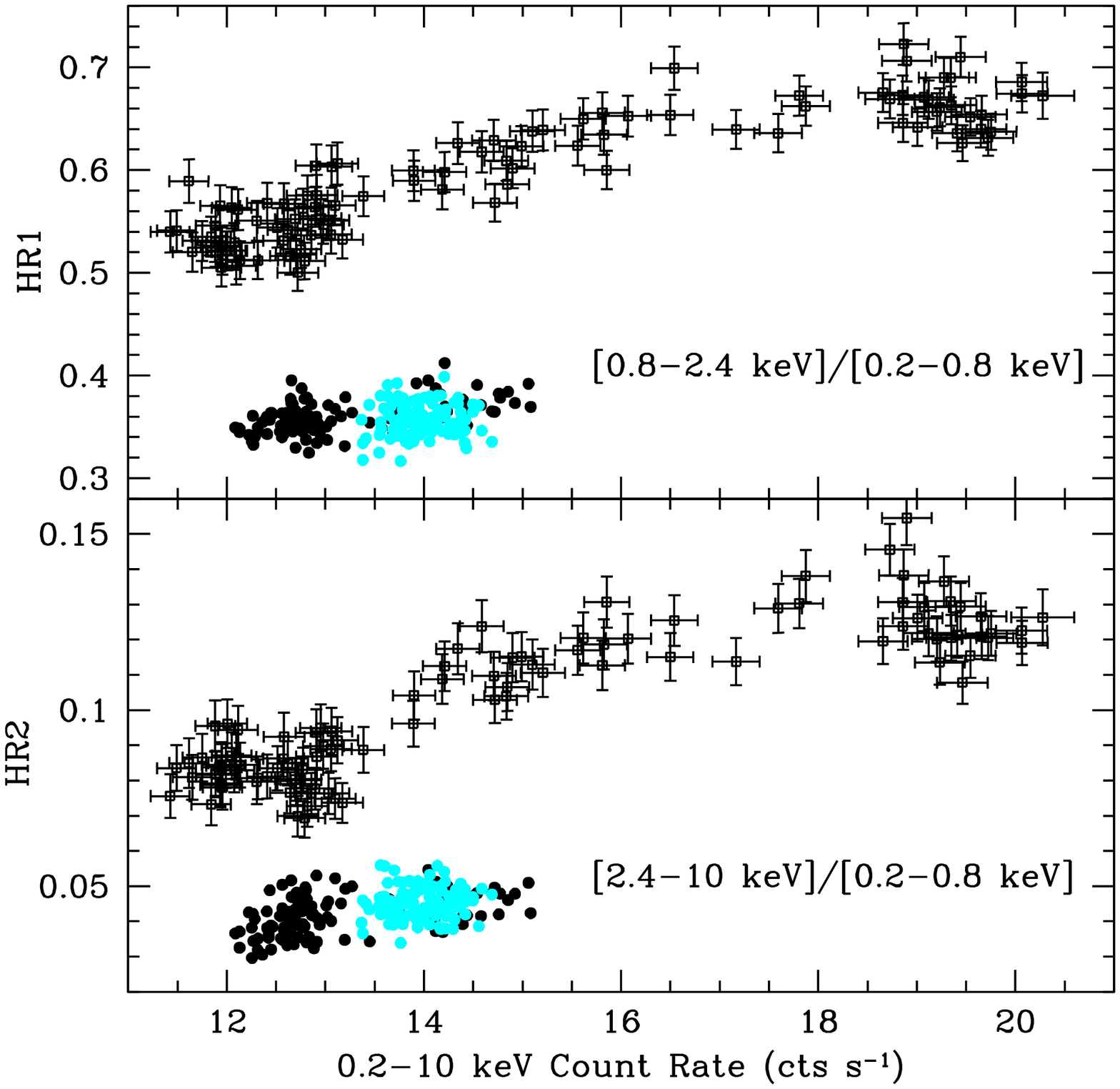}
\caption { \footnotesize As Figure~\ref{fig:lc:174} but for the Orbit 450 observation. In the left picture, the offsets of the normalized light curves (upper panel) between the exposure 450-1 (medium filter), 450-2 (thin filter) and 450-3 (thick filter) are caused by the different normalization with their own mean count rates, respectively. The different filters used also result in the offsets in the hardness ratios between the three exposures as seen from the medium and bottom panels, which are particularly significant between the exposure 450-2 ans 450-3.
In the right picture, the error bars for the exposure 450-1 (dark) and 450-2 (grey) are not shown for clarity, which are similar to those of the exposure 450-3 (dark with error bars). Also note that the different filters adopted give offsets in the total 0.2--10~keV count rates between the three exposures. }
\label{fig:lc:450}
\end{figure}

\begin{figure}
\plottwo{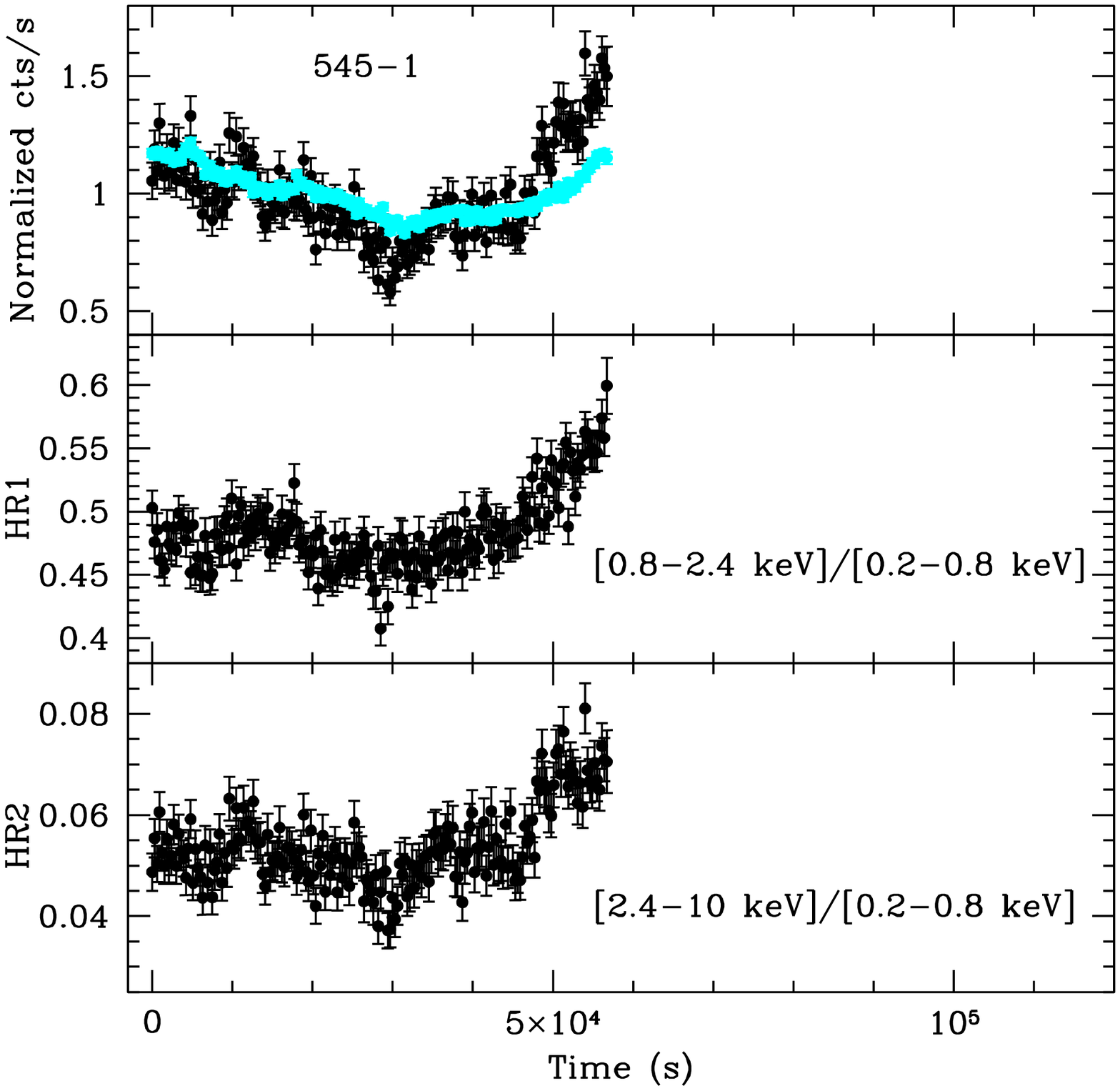}{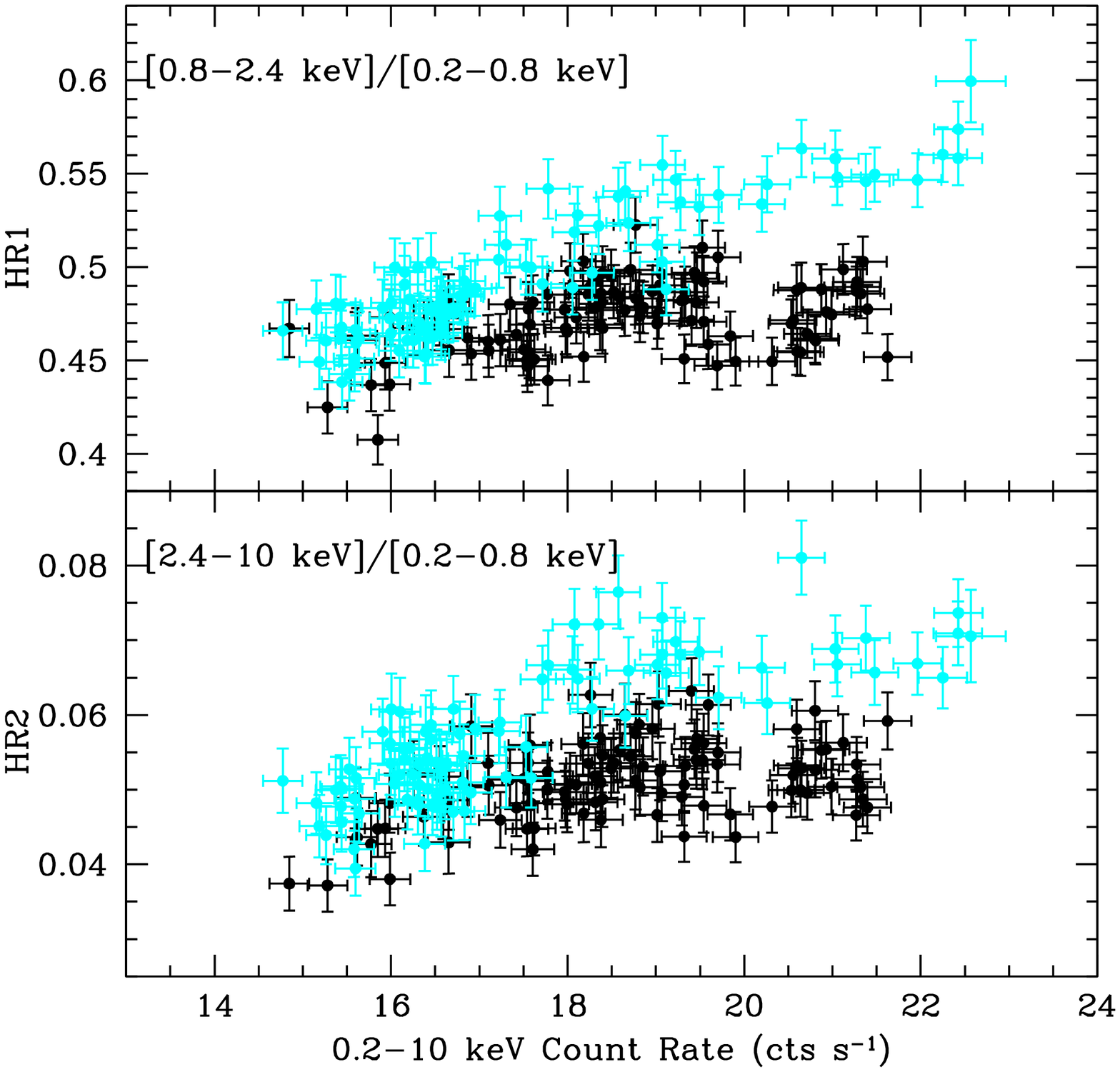}
\caption { \footnotesize As Figure~\ref{fig:lc:174} but for the Orbit 545 observation. In the right picture, the relationships between the hardness ratios and the total 0.2--10~keV count rates are plotted with dark and grey color for the decaying and rising part of the light curve, respectively. It is clear that the spectral variability rate is larger for the rising part than for the decaying part. }
\label{fig:lc:545}
\end{figure}

\begin{figure}
\plottwo{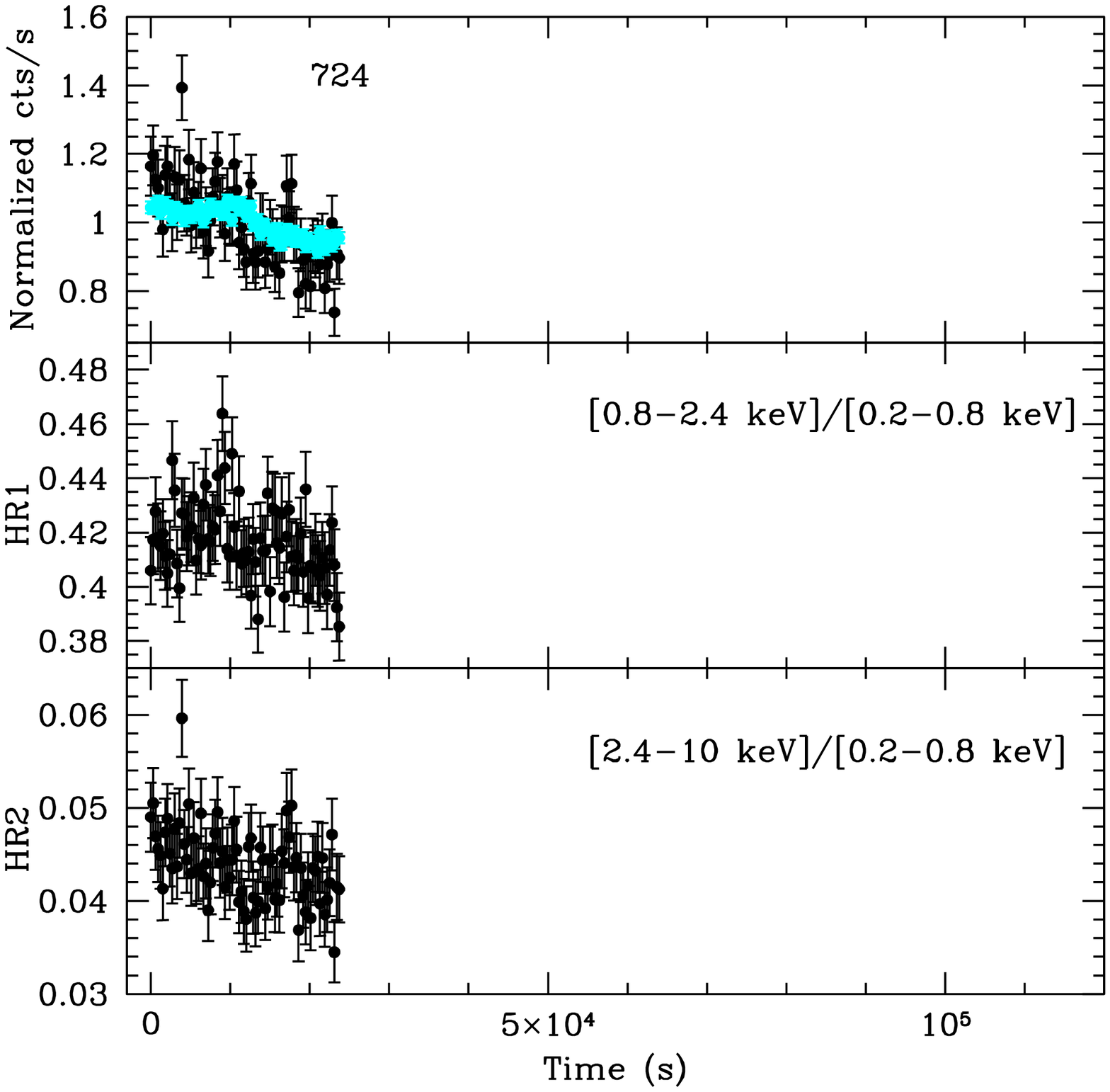}{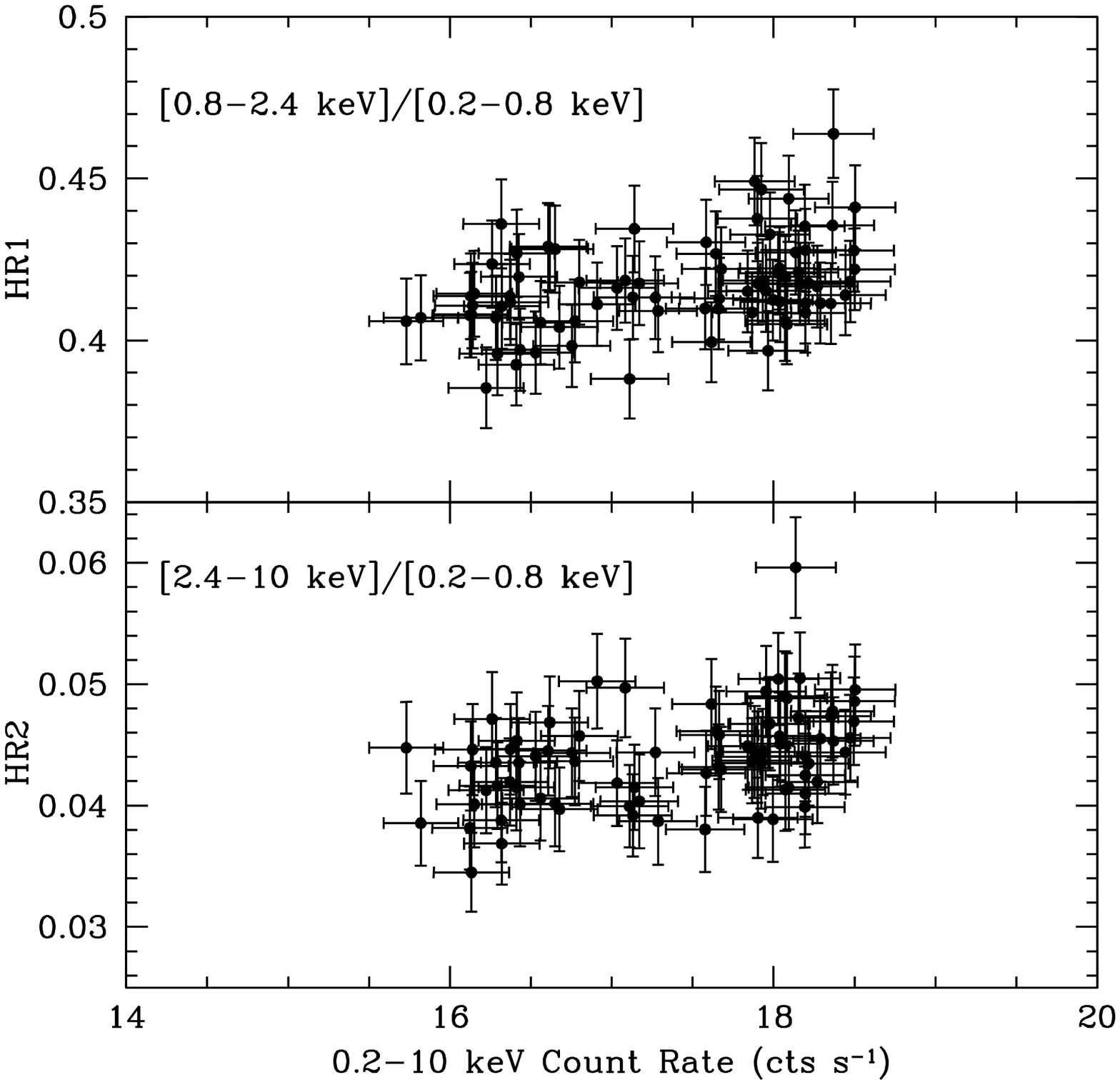}
\caption { \footnotesize  As Figure~\ref{fig:lc:174} but for the Orbit 724 observation.}
\label{fig:lc:724}
\end{figure}

The broad 0.2--10~keV energy band light curves have been presented in Paper I. Here we extracted, for each observation or exposure, the background subtracted light curves binned in 300~s in three energy bands, i.e., the 0.2--0.8~keV (soft) band, the 0.8--2.4~keV (medium) band and the 2.4--10~keV (hard) band. The soft (grey, cyan in color ) and hard (dark) band light curves, normalized by their mean count rates, respectively, are plotted together in the top panels of the left pictures of Figure~\ref{fig:lc:174}-\ref{fig:lc:724}. Light curves normalized in this way and plotted together provide a visual method for comparing the variability amplitudes, the correlations and possible time lags of the variability between the different energy bands. In all figures, we use the same scale for both the normalized count rate axis and the time axis in order for directly comparing the variability behavior (e.g., amplitude, timescale) of the source from epoch to epoch.     

Hardness ratio (HR) provides a model-independent method to study the spectral variability of a source. We calculate two hardness ratios: HR1=[0.8--2.4~keV]/[0.2--0.8~keV] and HR2=[2.4--10~keV]/[0.2--0.8~keV]. The evolutions of HR1 and HR2 with time are presented in the middle and bottom panels of the left pictures of Figure~\ref{fig:lc:174}-\ref{fig:lc:724}, respectively. One can see that the evolutions of the hardness ratios with time generally follows the light curves well. The right pictures of Figure~\ref{fig:lc:174}-\ref{fig:lc:724} plot the hardness ratios as a function of the total 0.2--10~keV count rates (top panels for HR1 and bottom panels for HR2), showing the relationship between the hardness ratios and the count rates. However, different extraction areas, different positions of the source on the chips, different pn configurations (imaging/timing mode with thin/thick/medium filter) can cause offsets of the light curves and the hardness ratios collected during different exposures, so we do not make a direct comparison of the light curves and hardness ratios between the different exposures of the same orbit and between the observations of different orbits.


\subsection{Orbit 174}

The top panel of the left picture of Figure~\ref{fig:lc:174} plots the normalized soft (grey) and hard (dark) band light curves toghether for the orbit 174 observation. The gap between the end of the exposure 174-1 and the start of the exposure 174-2 is $\sim 8380$~s, and the gap inside the exposure 174-2 are due to high particle background. The offsets of the normalized (soft and hard band) light curves between the exposure 174-1 and 174-2 are caused by the different normalization by their own average count rates, respectively. For the exposure 174-1, the variability amplitude is larger in the hard light curve than in the soft light curve since the overall slope of the flux decrease in the hard light curve is steeper than that in the soft light curve. The fractional $rms$ variability amplitude is $\sim 18$\% and $\sim 8$\% in the hard and soft band light curves, respectively.

Because of the small energy dependence of the pn detector over the point spread function (PSF) on the chips and the same filter (Thin) used, the hardness ratios of the two exposures can be directly compared. Therefore, the medium and bottom panels of the left picture of Figure~\ref{fig:lc:174} do not show offsets of hardness ratios between the two exposures. Both HR1 and HR2 variations follow the flux variations for the exposure 174-1, indicating flux-related spectral changes during the long flux decaying trend: the lower the fluxes, the softer the spectra. This is supported by the positive correlations between the hardness ratios and the count rates for the exposure 174-1, which are shown in the right picture of Figure~\ref{fig:lc:174} that plots the hardness ratios as a function of the total 0.2--10~keV count rates. This correlation, however, is not clear for the exposure 174-2 due to marginal variations.  

\subsection{Orbit 362} 

During the orbit 362 observation, the exposure 362-1 was operated in the timing mode while the exposure 362-2 was in Small Window imaging mode. Both exposures used the Medium filter. The gap between the two exposures is $\sim 1417$~s. The exposure 362-1 caught a strong flare, but the flare rising phase was possibly not fully sampled. The long decay phase ($\sim 40$~ks) of the flare may continue till the end of the exposure 362-2 (see Figure 2 of Paper I, note that the light curve of the exposure 362-1 was scaled down). The left picture of Figure~\ref{fig:lc:362} plots the normalized light curves in the soft (grey) and hard (dark) band (top panel), and the HR1 (medium panel) and HR2 (bottom panel) evolutions with time, respectively. As the orbit 174 observation, the offsets in the normalized light curves of the two exposures are due to the normalization by their own mean count rates, respectively. We can not make direct comparisons of hardness ratios between the two exposures because different operating modes, especially calibration uncertainties and increased noise below 0.5~keV in the timing mode, resulted in the significant offset of HR1 curves between them. The offset of HR2 curves is not significant.

During the flare 362-1, the hard light curve is significantly more variable than the soft light curve. The fractional $rms$ variability amplitude is $\sim 22$\% and $\sim 9$\% for the hard and soft band light curves, respectively. The variations of both HR1 and HR2 closely follow those of the light curves, indicating strong spectral evolution during the flare. The rigth picture of Figure~\ref{fig:lc:362} plots the hardness ratios as a function of the total 0.2-10~keV counts rates (dark) (note that the count rates have been arbitrarily scaled down by a factor of 4.75 for clarity), clearly showing positive correlations between the hardness ratios and the count rates: the spectrum softens as the flare flux decreases. However, the correlation of HR2 with the count rates is more complicated than that of HR1 with the count rates. At the lowest fluxes, the values of HR2 appear to be constant with the count rates.  
 
During the exposure 362-2, a flicker-like event was observed in the middle of the exposure. The flicker is easily visible in the hard light curve, and is also indicated by the hardness ratio curves. Both HR1 and HR2 curves follow the hard band light curve. The relationships between the hardness ratios and the count rates are shown in the right picture of Figure~\ref{fig:lc:362} (grey). Positive correlation is visible between HR1 and the count rate.

\subsection{Orbit 450}

The orbit 450 observation consists of three exposures, each lasting about 30~ks. The gaps between them are $\sim 803$~s and $\sim 1404$~s, respectively. The left picture of Figure~\ref{fig:lc:450} presents the normalized hard (dark) and soft (grey) band light curves (top panel), and the variations of HR1 (medium panel) and HR2 (bottom panel) with time, respectively. As the orbit 174 and 362 observations, the offsets in the normalized light curves of the three exposures are due to the normalization by their own mean count rates, respectively. The right picture of Figure~\ref{fig:lc:450} plots HR1 and HR2 as a function of the total 0.2--10~keV count rates. The observations were operated in Small Window imaging mode with different filters: 450-1 in Medium filter, 450-2 in Thin filter, and 450-3 in Thick filter. Because the thin filter is more transparent to the soft energy photons, the HR values can not be directly compared between the three exposures.  

A flicker occurred at the start of the exposure 450-1, it is stronger in the hard band than in the soft band. This weak feature was also followed by the variations of HR1 and HR2. In the HR-count ratio plot the correlations between the hardness ratios and the count rates is also visible. For the exposure 450-2, both the light curves and the hardness ratios are almost constant with time, and therefore there are no significant correlations between the hardness ratios and the count rates. 

The most violent event during this orbit observation is the strong flare during the exposure 450-3. The rising part of the flare was fully sampled while the decaying part was possibly not sampled at all. As in the flare 362-1, similar variations but different amplitudes were observed in the hard and soft light curves. The fractional $rms$ variability amplitude is $\sim 34$\% and $\sim 15$\% for the hard and soft band, respectively. The variations of the hardness ratios closely follow those of the hard band light curve, in agreement with the hardness ratio-count rate plots showing the harder spectra with the higher fluxes.

As shown in Paper I, the light curves of the three exposures during the orbit 450 observation are in fact continuous except for the small gaps due to the slight re-orientation of the instrumental setups of changing filters. Different filters resulted in offsets of the count rates and the hardness ratios between the three exposures (different positions of the source on the chips and different extraction areas of the source photons also give rise to such offsets with relatively small amplitudes). The significant offsets of HR1 and HR2 curves between the exposure 450-2 and 450-3 (the left picture of Figure~\ref{fig:lc:450}) indicate that there is an increase of about 0.15 in HR1 and about 0.02 in HR2 for the exposure 450-3 because the thick filter blocked more low energy photons than the thin and medium filters do. The invisible offsets of HR curves between 450-1 and 450-2 means that the hardness ratios obtained with the medium filter are similar to those obtained with the thin filter. In the hardness ratio-count rate plots, the 450-1 and 450-2 data will move up to the range occupied by the 450-3 data after adding about 0.15 to HR1 and about 0.02 to HR2, respectively.   

\subsection{Orbit 545}

The orbit 545 observation includes two exposures. The exposure 545-1 was operated in Small Window imaging mode with thick filter, and the exposure 545-2 in timing mode with thick filter. The exposure 545-2 data are not presented here due to possible calibration uncertainties (see also Paper I).
 
The exposure 545-1 may be described as an ``anti-flare''. The rate of flux increase is significantly faster than that of flux decrease. The normalized hard (dark) and soft (grey) band light curves are plotted in the top panel of the left picture of Figure~\ref{fig:lc:545}, and the evolutions of HR1 and HR2 with time in the medium and bottom panels, respectively. The right picture of Figure~\ref{fig:lc:450} plots HR1 and HR2 as a function of the total 0.2--10~keV count rates. The variability amplitude is larger in the hard than in the soft light curve, especially in the rising part of the normalized light curves. The fractional $rms$ amplitude is $\sim 18$\% and $\sim 9$\% for the hard and soft light curves, respectively. The variations of both HR1 and HR2 closely follow those of the normalized ligth curves. However, the correlations between the hardness ratios and the total count rates are not monotonous, and the spectral variability rate of the rising phase (grey) is significantly larger than that of the decaying phase (dark).

\subsection{Orbit 724}

Figure~\ref{fig:lc:724} shows the normalized hard (dark) and soft (grey) band light curves and the variations of HR1 and HR2 with time (left picture), and the HR1 and HR2 as a function of the total 0.2--10~keV count rates (right picture), respectively. The correlations between the hardness ratios and the count rates are visible.


\section{Cross-correlation Function Analysis}\label{sec:ccf}

\begin{figure}
\plottwo{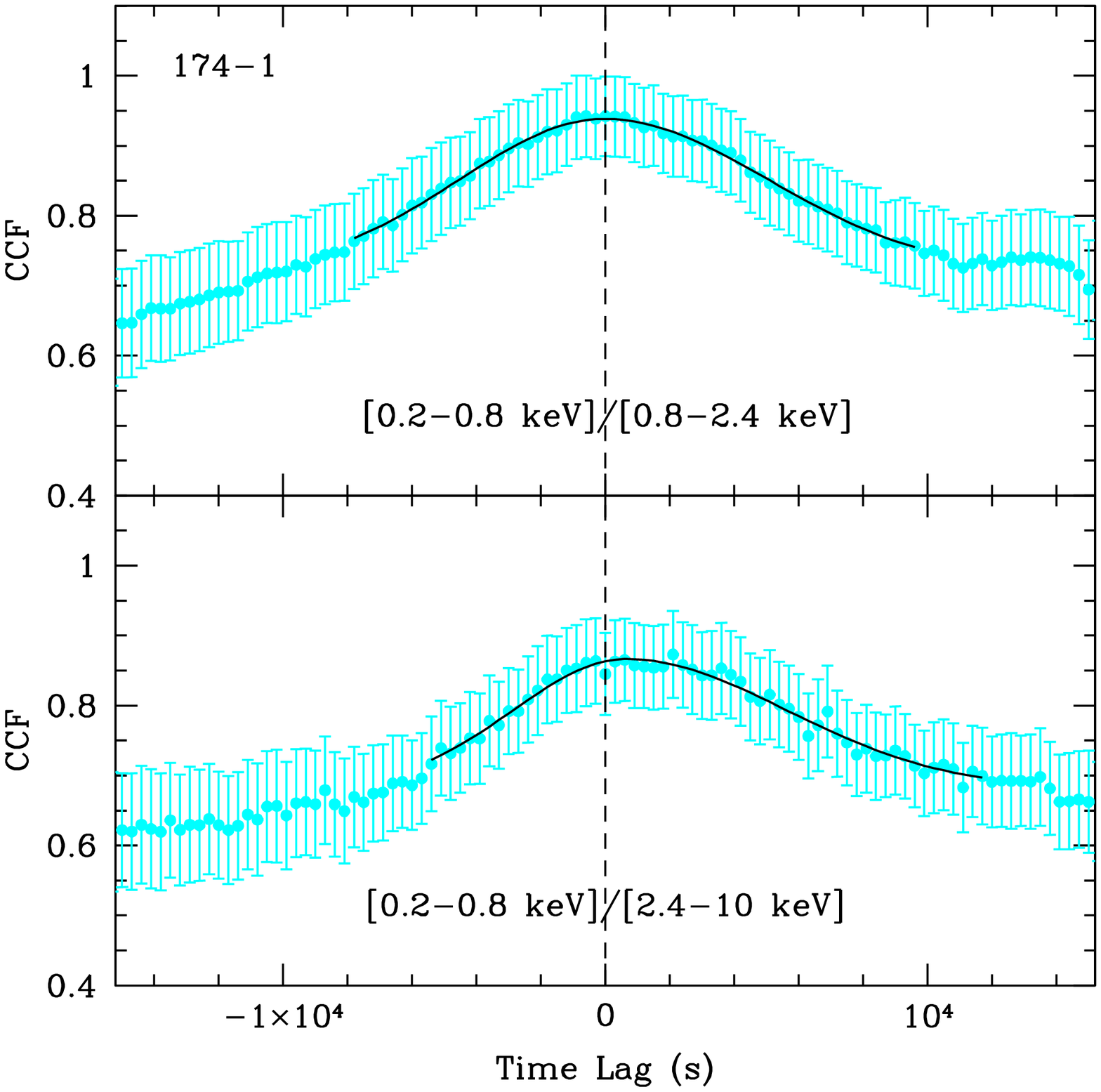}{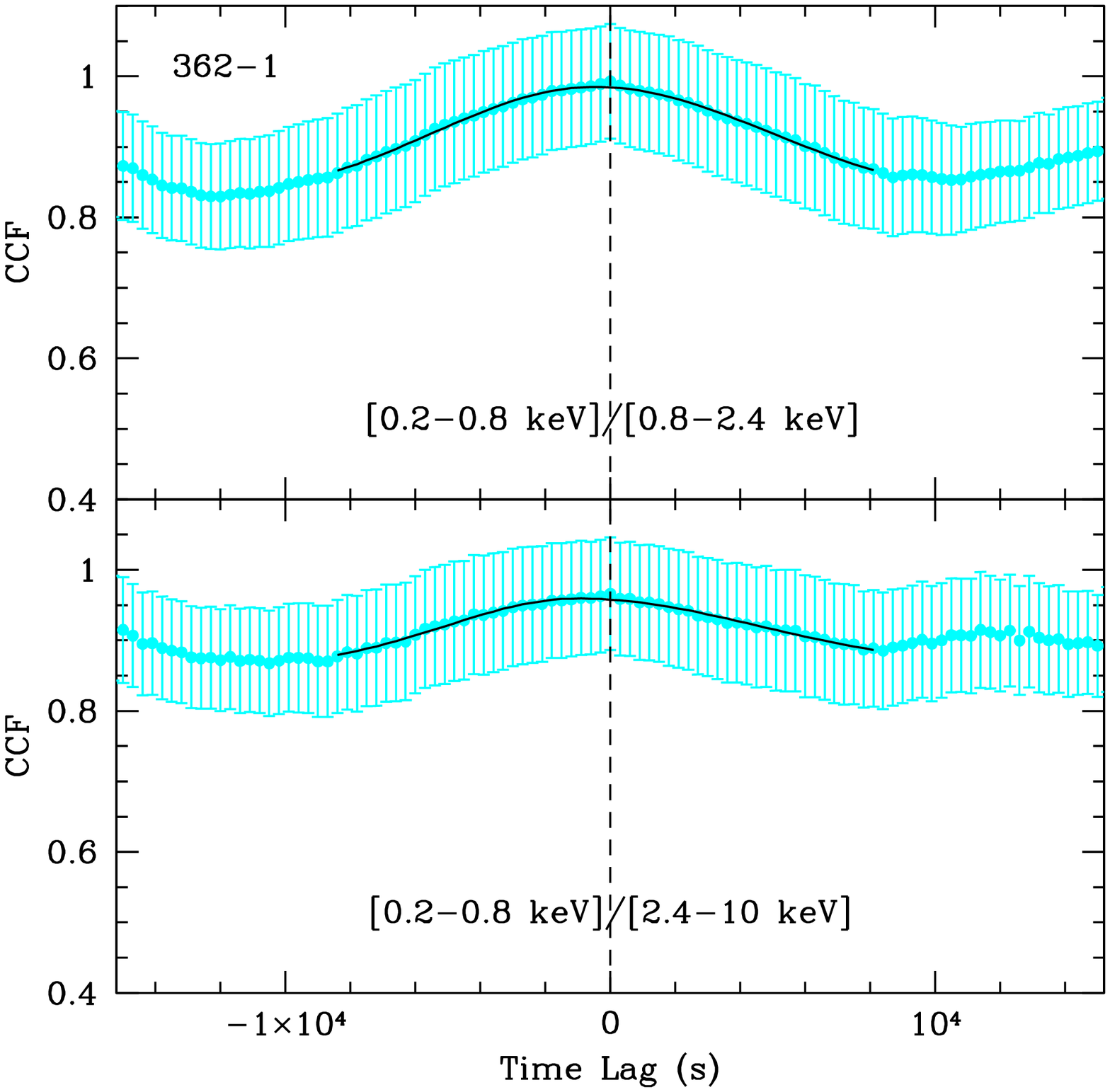}
\caption { \footnotesize The central $\pm 15$~ks part of the CCF (points with error bars in grey color). The solid black line is the best-fit with an asymmetric Gaussian function plus a constant. Left plot is for the exposure 174-1, and right plot for the exposure 362-1. }
\label{fig:ccf:174}
\end{figure}

\begin{figure}
\plottwo{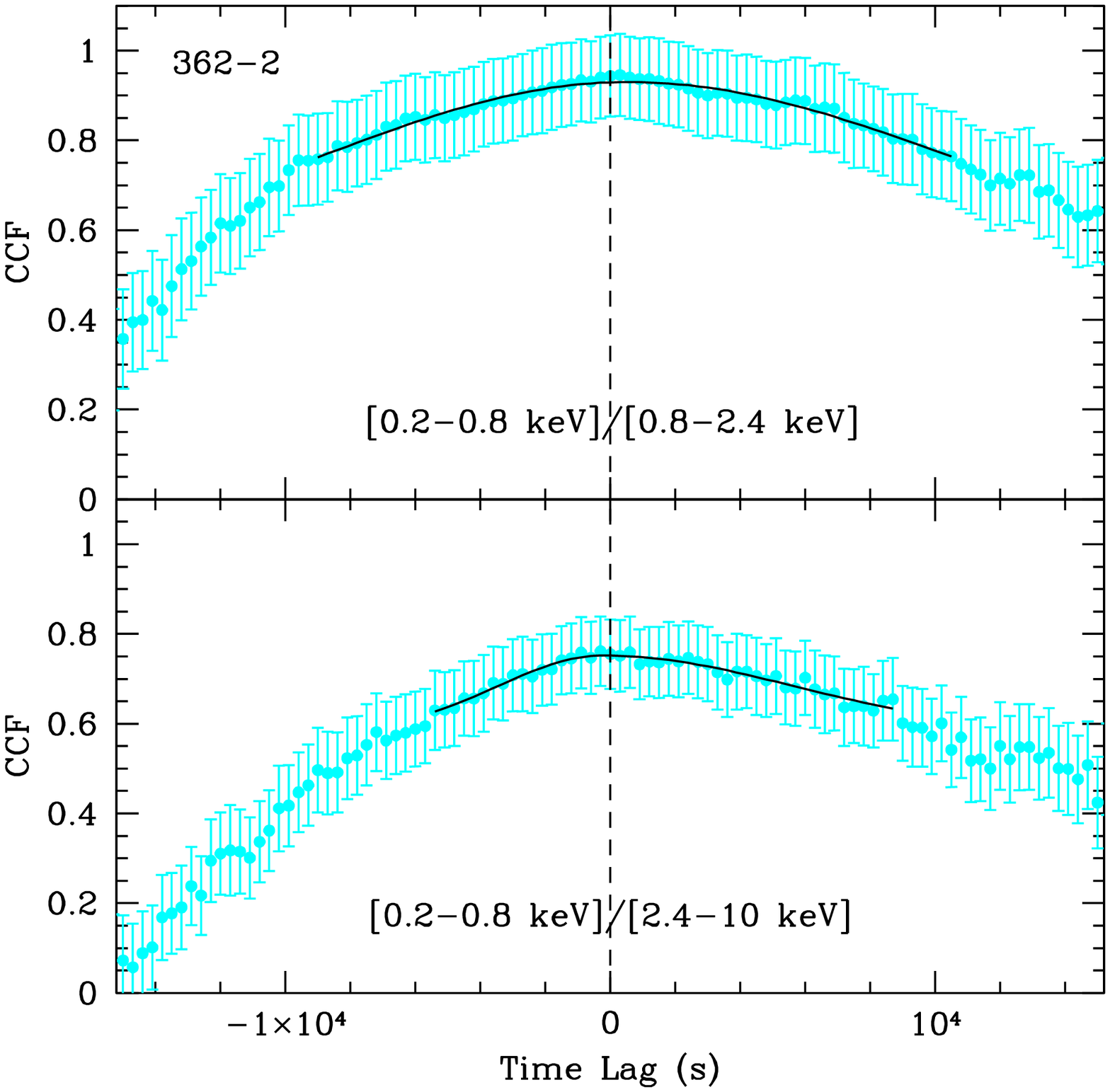}{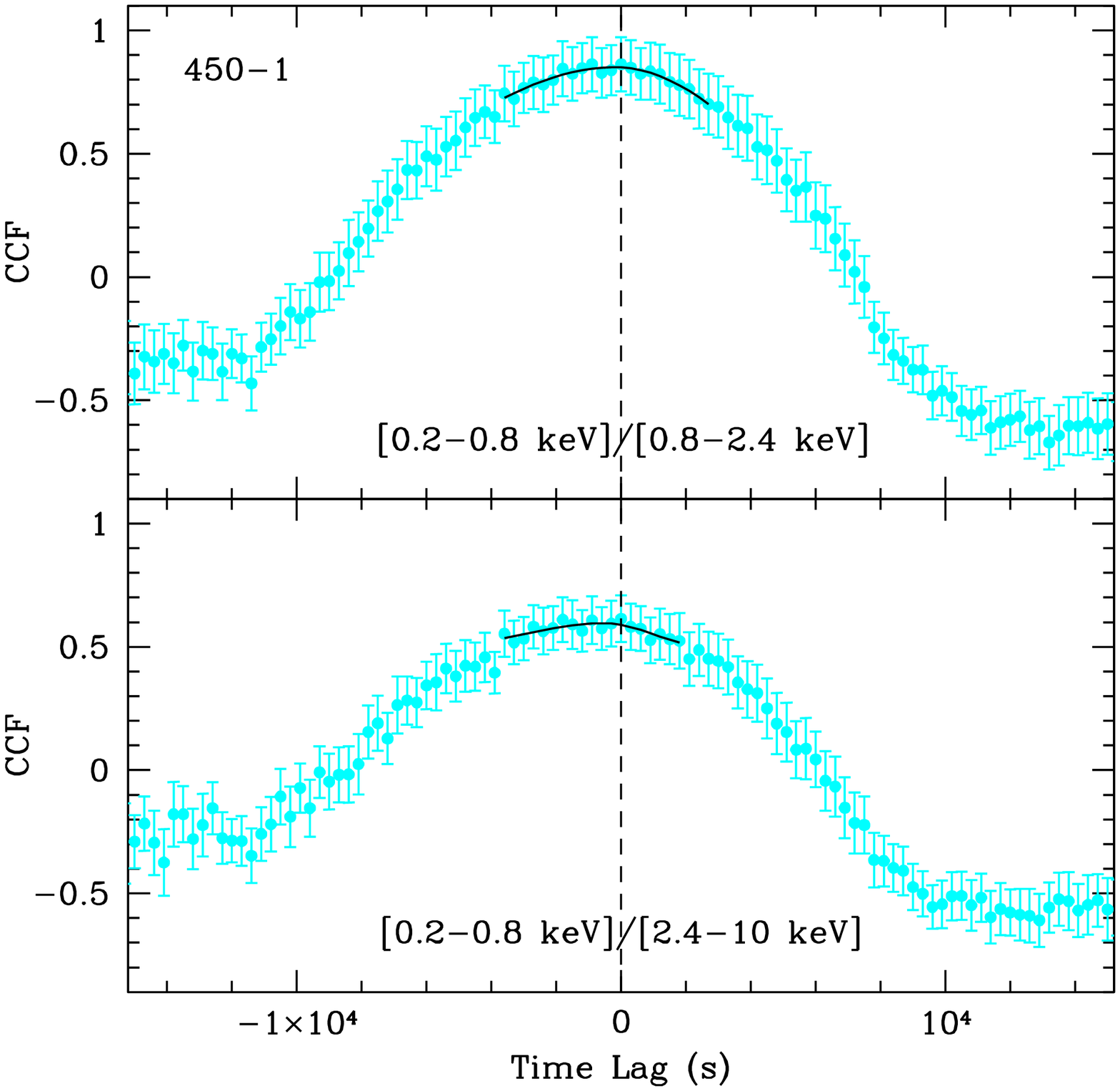}
\caption { \footnotesize As Figure~\ref{fig:ccf:174} but for the exposures 362-2 (left plot) and 450-1 (right plot). }
\label{fig:ccf:362}
\end{figure}

\begin{figure}
\plottwo{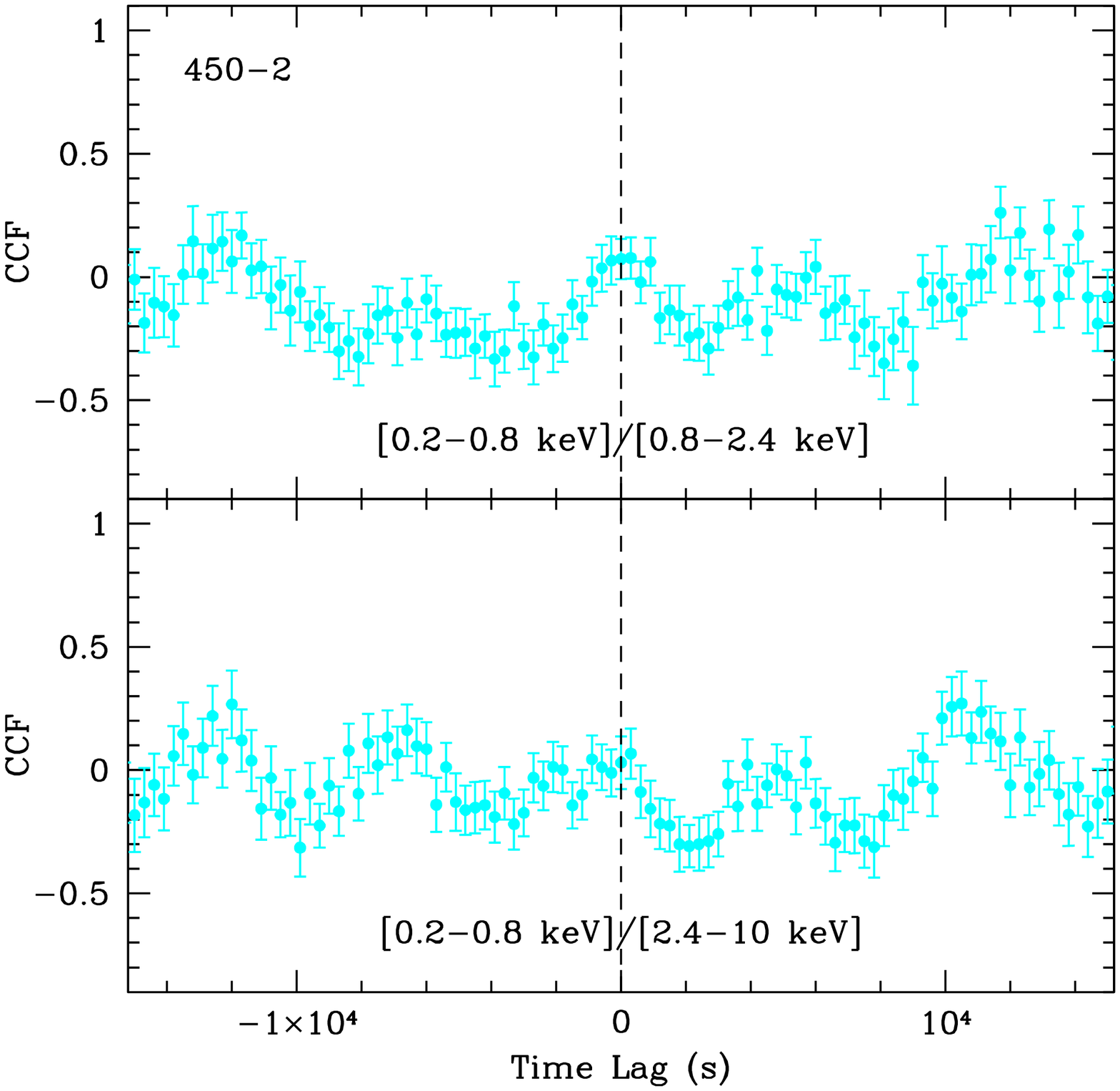}{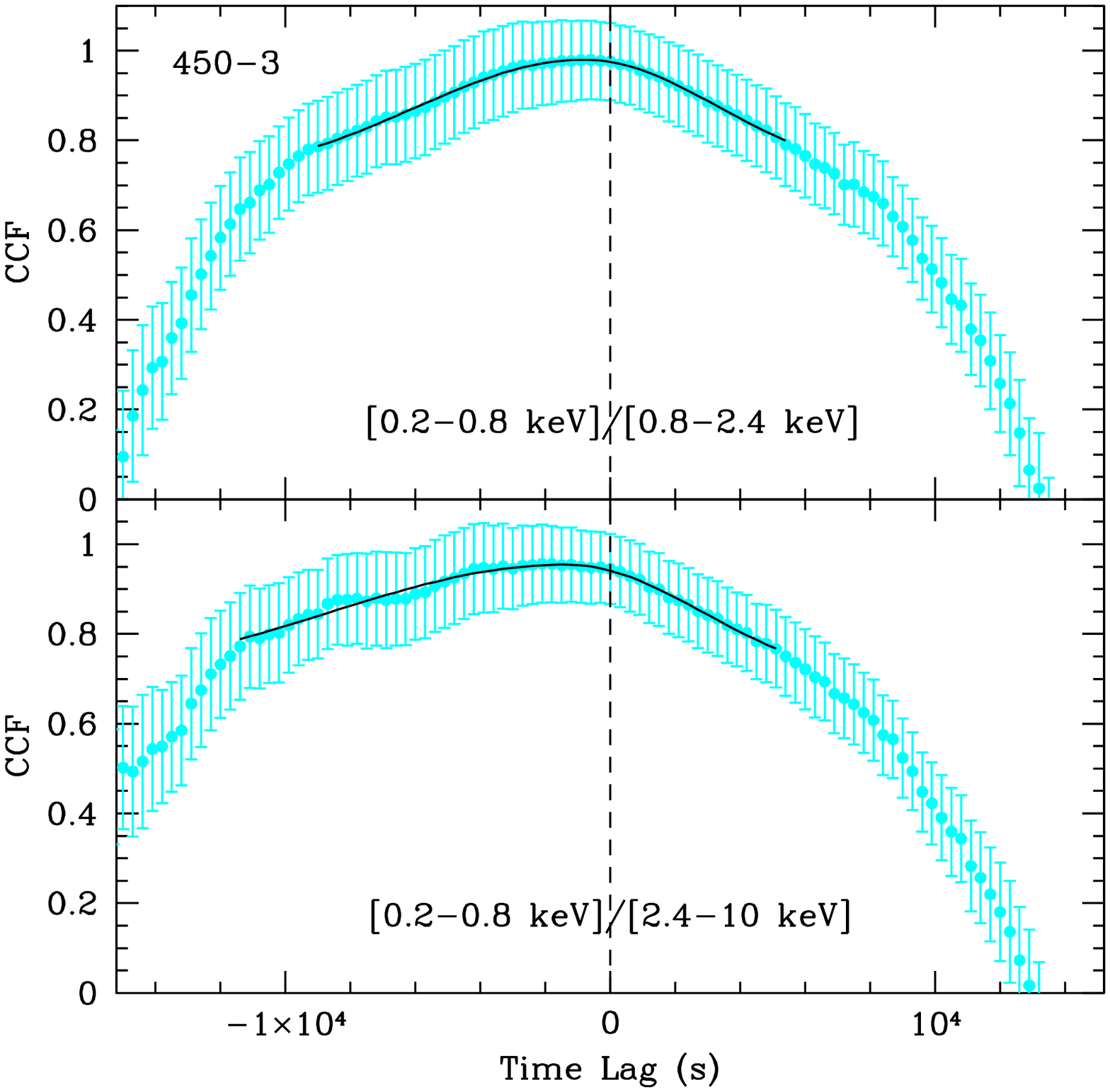}
\caption { \footnotesize  As Figure~\ref{fig:ccf:174} but for the exposures 450-2 (left plot) and 450-3 (right plot). The lags are not measured for the exposure 450-2 because of no correlations.}
\label{fig:ccf:450}
\end{figure}

\begin{figure}
\plottwo{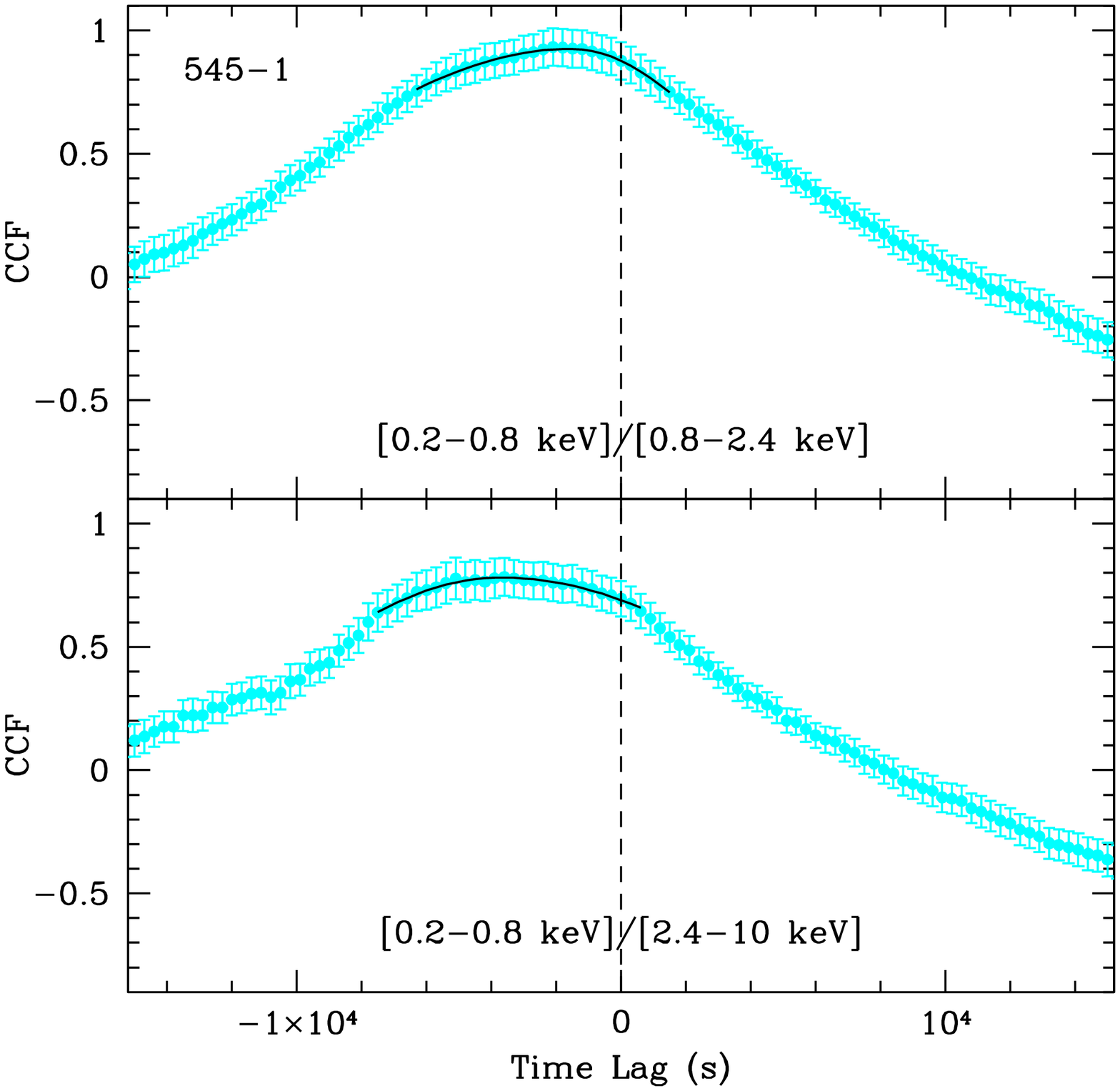}{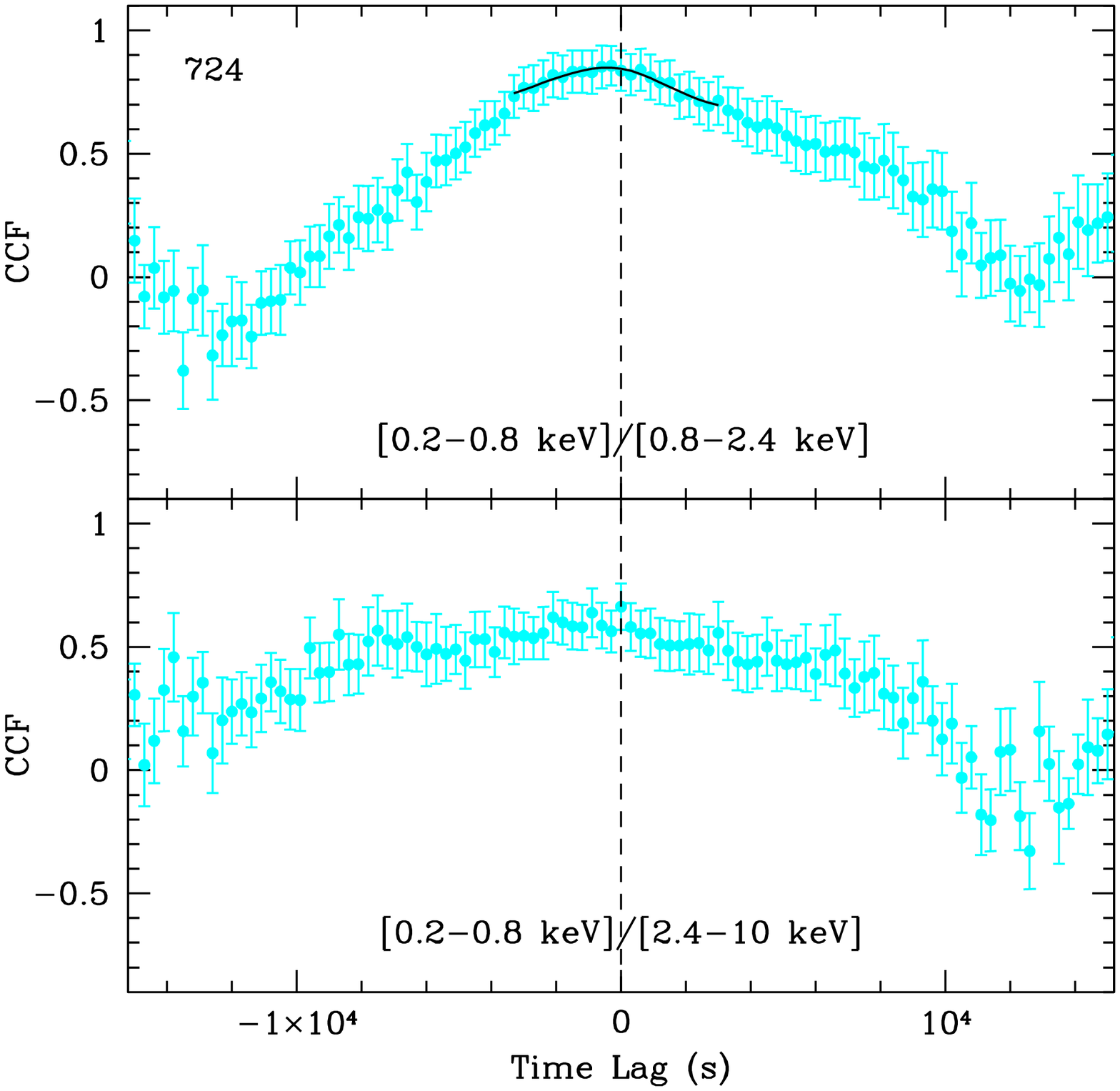}
\caption { \footnotesize As Figure~\ref{fig:ccf:174} but for the exposures 545-1 (left plot) and 724 (right plot). The lag is not measured for the soft/hard case of the exposure 724 because the CCF peak is not well defined.}
\label{fig:ccf:545}
\end{figure}

Because the light curves gathered with EPIC pn detector are uniform (i.e., without gaps), we can calculate the standard Cross-Correlation Function (CCF) between the light curves in the different energy bands. Paper I showed that the light curves of \pks\ are not stationary, so the CCF must be normalized by the mean count rates and the standard deviation of the two cross-correlated light curves using only the data points that actually contribute to the calculation of the CCF at each lag (White \& Peterson 1994). With the same 300~s binned light curves used in \S\ref{sec:hr}, we compute CCFs in two cases: 0.2--0.8~keV versus 0.8--2.4~keV (i.e., soft vs medium energy band) and 0.2--0.8~keV versus 2.4--10~keV (i.e., soft vs hard energy band). The results are presented in Figures~\ref{fig:ccf:174}-\ref{fig:ccf:545} (points with error bars in grey colour). For clarity, we plot only the central $\pm 15$~ks lag range of each CCF. One can see that the CCF shapes of the source are very complicated, and clearly different from epoch to epoch. The solid black line superimposed on each CCF represents the best-fit model with an asymmetric Gaussian function plus a constant (see below). 

We measure the inter-band time lags with three techniques: (1) searching for the CCF peak value ($r_{\rm max}$), the lag corresponding to the peak is marked as $\tau_{\rm peak}$; (2) computing the CCF centroid over lags bracketing $r_{\rm max}$ (Peterson \et 1998), we use $\tau_{\rm cent}$ to represent the lag derived with this method; (3) fitting the CCF with an asymmetric Gaussian function (plus a constant), the lag associated with the peak of the function is called $\tau_{\rm fit}$ (Brinkmann \et 2003; Ravasio \et 2004). After finding $r_{\rm max}$, we measure the time lag with method (2) and (3) only using the CCF points with $r$ in excess of 0.8$r_{\rm max}$ (Peterson \et 1998). The errors on the lags, estimated with the Monte Carlo method suggested by Peterson \et (1998), represent 68\% confidence limits around the median values of the simulation results. A positive lag associated with strong correlation indicates that the lower energy variations lead ahead the higher energy (i.e., hard lag), and a negative lag associated with strong correlation indicates the opposite case: the lower energy variations lag behind the higher energy (i.e., soft lag). Table~\ref{tab:lag} tabulates the lags measured with different methods and the corresponding simulation results.

There are a number of caveats for the lag measurements, mainly relating with the complexities of the CCFs. If one increases or decreases the time lag range of the CCF used to estimate the lag, different lags may be obtained with both $\tau_{\rm cent}$ and $\tau_{\rm fit}$ methods. The CCF complexities may also result in different lags if one measures the lags with different methods on the basis of the same CCF lag range. It appears that $\tau_{\rm cent}$ accounts for the CCF asymmetries better than \tfit\ does (\tpeak\ do not account for the CCF asymmetries at all). If the CCFs were smooth and symmetric functions, the lags measured with different methods should be consistent with each other, and would not significantly depend on the CCF range selected. It is worth noting that the time lags measured with different techniques are usually inconsistent with each other, and in some cases the differences are quite large. We found that the lags measured with \tcent\ method are usually larger than those measured with \tpeak\ and \tfit\ in most cases (see Table~\ref{tab:lag}). The significant CCF complexities (such as asymmetries and irregularities) may yield such inconsistence. Asymmetric function (better than Gaussian function) may account for the CCF asymmetries but not the CCF irregularities. Perhaps \tcent\ is the best technique to measure lags with complicated CCFs. The CCF complexities may be caused by the complexities of the light curves, i.e., an representation of the intrinsic complexities of the underlying variability processes. The CCF complexities may not be associated with the photon statistics.

The left picture of Figure~\ref{fig:ccf:174} plots the two CCFs for the exposure 174-1. We find that the light curves at different energy bands are strongly correlated, which is also visible from the normalized light curves as shown in Figure~\ref{fig:lc:174}. The CCF peak values (\rmax) are 0.94 and 0.87 for the soft/medium and soft/hard CCFs, respectively, indicating that the soft/medium band light curves are more correlated than the soft/hard light curves. Note that the CCFs for this exposure represent the correlations only for the long decaying trend variability. The soft/medium CCF appears to be symmetric near zero lag, with slight asymmetries toward the positive lags. The simulations confirm this feature: \tpeak\ and \tfit\ are consistent with zero lag, but \tcent\ suggests a small positive lag. In contrast, the soft/hard CCF shows larger asymmetries toward positive lags than the soft/medium CCF does, confirmed by the simulation results that yield hard lags with higher confidence compared to the previous cases. From the point view of \tcent, the higher energy variations are lagging behind the lower energy, and the lags are larger with larger differences between the energy range considered. We do not calculate CCF for the exposure 174-2 due to the long gap. 

In the right picture of Figure~\ref{fig:ccf:174} we plot the two CCFs for the exposure 362-1, a strong flare though the rising part was not fully sampled. The two CCFs show strong peaks around zero lag, and appears to be symmetric. The maximum correlation coefficients of both the soft/medium and soft/hard CCF are $\sim$ 1 (the soft/medium \rmax\ is 0.99, and the soft/hard \rmax\ 0.97), indicating that the variability between the different energy bands is well correlated. The simulations suggest that \tpeak\ is consistent with zero lags, and \tcent\ $\sim -2.5$ min for both the the soft/medium and soft/hard. \tfit\ is $\sim -7.5\pm2.5$~min and $\sim -15\pm4$~min for the soft/medium and the soft/hard, respectively, indicating that the soft lag is larger with larger energy difference.

The two CCFs for the exposure 362-2 are plotted in the left picture of Figure~\ref{fig:ccf:362}, they show very broad peaks. \rmax\ is 0.95 and 0.76 for the soft/medium and soft/hard CCF, respectively, indicating that the correlation between the soft and hard band is weaker than that between the soft and medium band. The soft/hard CCF is more asymmetric toward positive lags than the soft/medium one. The simulations with \tpeak\ method gives rise to lags consistent with zero. However, the simulations with \tcent\ suggest that the lower energy variations lead ahead the higher energy. 

The exposure 450-1 showed a flicker-like event at the beginning of the exposure. The right picture of Figure~\ref{fig:ccf:362} plots the two CCFs. \rmax\ is smaller than the previous cases, which is 0.86 and 0.76 for the soft/medium and soft/hard CCF, respectively. Weaker correlations may be expected from smaller amplitude of variability. It appears that the CCFs are symmetric near the zero lag. The simulations with \tcent\ method suggest that the soft band variations lag behind the medium and hard band by $\sim 7.5\pm5$~min and $\sim 14.5\pm9$~min, respectively.

The exposure 450-2 did not show significant variability. The left picture of Figure~\ref{fig:ccf:450} shows the two CCFs, indicating that the light curves in the different energy bands are not correlated, mainly caused by the lack of flux variations.

The exposure 450-3 revealed a strong flare. The two CCFs are presented in the right picture of Figure~\ref{fig:ccf:450}, they are broad and asymmetric toward negative lags, and clearly peak at negative lags. The large values of \rmax\ (0.98 and 0.96 for the soft/medium and soft/hard CCF, respectively) indicate that the variations in the different energy bands are well correlated. The simulations with different methods all suggest that the soft band variations are delayed with respective to the medium and hard bands: \tpeak\ is 15 and 30~min, \tcent\ $\sim 27\pm5$ and $\sim 47\pm7$~min, \tfit\ $\sim 14\pm3$ and $\sim 23\pm6$~min for the soft/medium and the soft/hard bands, respectively.
 
The two CCFs for the exposure 545-1, plotted in the left picture of Figure~\ref{fig:ccf:545}, are asymmetric toward negative lags and clearly peak at negative lags. Therefore, the soft band variations lag both the medium and hard bands. The correlation between the soft and medium band (\rmax\ $\sim 0.93$) is stronger than that between the soft and hard band (\rmax\ $\sim 0.78$). The simulations with different methods give soft lags of about half hour and one hour for the soft/medium and soft/hard CCF, respectively. These are the maximum lags found among all the exposures presented in this work.  

The right picture of Figure~\ref{fig:ccf:545} presents the two CCFs for the exposure 724. The soft/medium CCF shows asymmetries toward negative lags in small lag range while it shows asymmetries toward positive lags in large lag range. The value of \rmax\ is 0.86. The soft/hard CCF is very broad without well defined peak, so we do not measure the lag in this case. The simulations (\tcent) suggest that the soft band variations lag behind the medium band by about 5~min with large uncertainties.

Table~\ref{tab:lag} shows that different methods generally do not yield consistent lags in most cases, so we calculate, with the simulation values, the probability that a lag is detected as either negative (soft lag), positive (hard lag) or zero lag. The probabilities which are larger than $95\%$ are considered as significant. The maximum probability for each lag is shown in Table~\ref{tab:prob}. The results clearly show that a significant detection of a lag is sensitive to the presence of a peak or trough in the light curves. Significant soft lags were detected in the exposures 362-1, 450-3 and 545-1. For the exposure 362-1, however, \tpeak\ is zero with $\sim 100\%$ probability, caused by the fact that the lags, as seen from \tcent, may be smaller than the binsize (300~s) adopted. Moreover, \tpeak\ does not take into account the CCF asymmetries at all. For other exposures without strong peaks in the light curves, the detections of lags are significant only for \tcent. Finally, it is interesting to note that the correlations of the variability in the different energy bands are identically strong for the two flares 362-1 and 450-3 (\rmax\ is $\sim 1$). However, in other cases the correlations between the soft and hard band are weaker than those between the soft and medium band. 

\section{Discussion}\label{sec:disc}

\subsection{The spectral variability}

In paper I, the rms spectra derived with the \xmm\ EPIC pn light curves showed that the fractional rms variability amplitude of \pks\ systematically increases with increasing photon energies, which is especially significant during active epochs (either presence of flares or strong flux rising/decreasing variability trends). This strongly indicates that the flux variations of the source must be accompanied by spectral variations as well, which is confirmed by the model-independent hardness ratio analysis of the 300~s binned light curves presented in \S\ref{sec:hr}. The X-ray spectra generally harden (i.e., hardness ratios increase) when the source fluxes increase, and vice versa. Similar behavior of the X-ray spectral variability was already observed in other X-ray observations of \pks\ (e.g., Zhang \et 2002), as well as in other TeV blazars (e.g., Mrk~421, Fossati \et 2000; Brinkmann \et 2003; Ravasio \et 2004; Mrk~501, Pian \et 1998; IES 2344+514, Giommi, Padovani \& Perlman 2000). However, it is worth noting that the spectral variations in the hard band are more complicated than in the soft band (e.g., the exposure 362-1).

\subsection{The inter-band time lags}

Using the same 300~s binned light curves as used in the hardness ratio analysis, we performed standard CCF analysis in order to search for possible inter-band time lags. Among the eight exposures, the inter-band light curves are well correlated in seven cases in which the source showed variations. The inter-band light curves are not correlated at all in the exposure 450-2 where the fluxes are least variable. The CCF peak correlation coefficients are stronger in flares than in other epochs. For the flares 362-1 and 450-3 (though they were not fully sampled), the soft band variations are well correlated with both the medium and hard bands. However, in other cases, the peak correlations between the soft and medium band are stronger than those between the soft and hard band, which may be caused by weak photon statistics in the hard band. Physically, the correlations of the variability in the different energy bands may change from epoch to epoch. 

Among the seven exposures in which the light curves are well correlated, the lower energy variations lag the higher energy in three cases with high significance, while in other four cases it is possible that the higher energy band variations lead or lag the lower energy band. From this and previous studies (e.g., Brinkmann \et 2003), it seems that the probability of detecting soft lags is larger than that of detecting hard lags. Regardless of the properties of the light curves (flare or not), it is worth emphasizing that the definite detection of soft lag of about 1 hour in the exposure 545-1, supports the significant lags detected with \sax\ and \asca. It also provides a specific case against the claim that these lags are artificial due to periodic gaps in the light curves gathered with low Earth orbital satellites (Edelson \et 2001). Therefore, the detections of lags do not depend on the sampling properties of the light curves. Using Monte Carlo methods, however, Zhang \et (2004) showed that the periodic gaps can increase the uncertainties of lag determinations.

We estimated the time lags on the basis of individual exposures, we did not computed CCFs with sub-intervals of the light curves because there are no multiple flares in the individual exposures. Previous studies showed that the lags can be significantly different when using different sub-parts of a light curve, mainly based on the presence of strong flares (e.g., Zhang \et 2002; Brinkmann \et 2003; Ravasio \et 2004). The light curves from which the lags were derived in this work have the following different properties: the exposures 362-1 and 450-3 are strong flares but not fully sampled, the exposures 174-1 and 362-2 are the long flux decaying trends, the exposure 545-1 is an ``anti-flare'',  and the exposures 450-1, 450-2 and 724 are absent of significant variability. If the ``flares'' were fully sampled, the time lags may be significantly different if one cross-correlates only the rise or decay part of the flares, respectively, instead of the whole flares. Orbit 362 observation demonstrates this proposition. If we believe that the exposure 362-2 is a continuity of the decaying phase of the flare 362-1, the CCF analysis gives different sign of lags: 362-1 showed soft lag but 362-2 possible hard lag. However, such an opposite behavior may be complicated by the superposition of a ``flicker'' on the 362-2 part of the whole decay phase of the flare. 

\subsection{Physical implications}

The inter-band time lags were frequently detected with various X-ray telescopes  during X-ray flares of \pks\ and Mrk~421. The origin of the lags is usually attributed to the energy-dependent acceleration and cooling timescales of the emitting particles. The sign of a lag (either soft or hard lag) is determined by the comparison of the two timescales. The cooling process is universally known, but the acceleration process is not well understood yet and could operate in different ways (e.g., Katarzinsky \et 2005). Following Zhang \et (2002), we assume that the diffusive shock acceleration is the mechanism of accelerating the electrons (e.g., Blandford \& Eichler [1987]), and synchrotron radiation is the cooling mechanism of the accelerated electrons. With these assumptions and in the observer's frame, the acceleration timescale \tacc\ and cooling timescale \tcool\ of the relativistic electrons can be expressed as a function of the observed photon energy $E$ (in keV):
\begin{equation}
t_{\rm acc}(E) = 9.65 \times 10^{-2} (1+z)^{3/2} \xi 
	B^{-3/2} \delta^{-3/2} E^{1/2} \quad {\rm s} \,,
\label{eq:tacc}
\end{equation}
\begin{equation} 
t_{\rm cool}(E) = 3.04 \times 10^{3} (1+z)^{1/2} B^{-3/2}
	\delta^{-1/2} E^{-1/2}  \quad {\rm s} \,,
\label{eq:tcool}
\end{equation}
where $z$ is the source's redshift, $B$ the magnetic field in Gauss and $\delta$ the Doppler factor of the emitting region, and $\xi$ the parameter describing how fast the electrons can be accelerated. Perhaps $\xi$ is the most important parameter in determining whether a flare behaves as soft or hard lag (see Zhang \et 2002 for details).
 
Equations (\ref{eq:tacc}) and (\ref{eq:tcool}) show that \tacc\ and \tcool\ have an inverse dependence on the observed photon energy: the lower energy electrons radiating the lower energy photons cool slower but accelerate faster than the higher energy electrons radiating the higher energy photons do. If \tacc\ is significantly shorter than \tcool, the cooling process dominates the system. The emission propagates from higher to lower energy, the higher energy photons then lead the lower energy ones (i.e., the so-called soft lag). On the HR-count rate (or spectral index-flux) plot, the spectral evolution follows a clockwise loop; If \tacc\ is comparable to \tcool, the acceleration process dominates the system. The emission propagates from lower to higher energy, so the lower energy photons lead the higher energy ones (i.e., so-called hard lag). If viewed from the HR-count rate plot, the spectral evolution tracks a counterclockwise loop. The observed fact that the variability of a source's flares shows soft lags in some epochs and hard lags in other epochs, implies that the difference between \tacc\ and \tcool\ of the emitting electrons is changing from flare to flare. Given $B$ and $\delta$ of the emitting region that radiate the observed 0.2-10~keV energy band studied here, \tcool\ is fixed. This further implies that \tacc\ must be changing with time. Equation (\ref{eq:tacc}) shows that the changes of \tacc\ are modulated by the acceleration parameter $\xi$. The internal shock scenario assumes that the relativistic outflow (jet) is inhomogeneous. The discrete blobs ejected by the central engine possess different initial velocities, and can collide each other at some distance from the center. The particle acceleration process takes place in the shock formation processes due to collisions (e.g., Spada \et 2001; Mimica \et 2005). So the changes of $\xi$ indicate that the shock formation mechanisms and/or the subsequent particle acceleration mechanisms may be not similar for all shock processes, or the ``shocked'' regions that will radiate soon-after, have different physical parameters. The consequence of changing $\xi$ (the acceleration rate) leads to the changes of the maximum energy ($\gamma_{\rm max}$) of the particles and the associated maximum photon energy ($E_{\rm max}$) of synchrotron emission. $E_{\rm max}$ is related with $\gamma_{\rm max}$ by the formula $E_{\rm max} = 3.7\times 10^6 \delta B\gamma_{\rm max}^2$~Hz \footnote{In fact, $E_{\rm max}$ is given by the balance between \tacc\ and \tcool\ in a single power law approximation of the electron distribution, but the situation is more complex, e.g., the electron distribution always has a ``break" (e.g., Kirk, Rieger, \& Mastichiadis 1998).}. 

The direct consequence of changing $\xi$ is that the energy difference of photons between the observed 0.2--10~keV and $E_{\rm max}$ is not similar at all time. If $E_{\rm max}$ is much higher than 10~keV, i.e., \tcool\ $\gg$ \tacc\ in the 0.2--10~keV band (since $t_{\rm acc}=t_{\rm cool}$ at $E_{\rm max}$), the cooling process dominates the system in the 0.2--10~keV band, and the higher energy photons will reach the observer first and the inter-band variability is characterized by soft lag. If $E_{\rm max}$ is similar to 10~keV, i.e.,  $t_{\rm cool} \sim t_{\rm acc}$ at 10~keV, \tacc\ dominates the system in the 0.2-10~keV band. In this case the variability propagates from lower to higher energy and the hard lag will be observed. The relative location of $E_{\rm max}$ on the SED of a source with respect to the 0.2--10~keV could be qualitatively inferred from the spectral index in the 0.2--10~keV: the softer (steeper) the spectra, the closer $E_{\rm max}$ relative to the 0.2--10~keV, and so \tcool\ is more comparable to \tacc\ in the 0.2--10~keV band. Then the inter-band variability may be observed as hard lag with larger probability. The two flares of Mrk~421 fully sampled by \xmm\ present an example to show the qualitative relationship between the lags and spectral indices. Ravasio \et (2004) presented both spectral indices and time lags of the two flares. Spectral indices ($\sim 1.55$) of the 2002 December 1 flare were significantly larger (softer) than those ($\sim 1.15)$) of the 2002 November 14 flare, which is qualitatively consistent with the detected lags: the former flare showed hard lag and the latter soft lag. This simple relationship between the two observed parameters should be further examined with more strong flares from different sources. In fact, the relationship between the sign of lag and the spectral index is a representation of the simulations by Kirk \et (1998). Their simulations, including both particle acceleration and synchrotron emission, showed the dependence of the sign of lag on the photon energies, i.e., soft lag was obtained in low energy band with flat spectral index, and hard lag in the high energy band with steep spectral index (Figures 3 and 4 of Kirk \et [1998]). Observationally, Zhang \et (2002) found, using the spectral index-flux plots, that the symmetric flare obtained with \sax\ in 1997 November showed soft lag in the soft X-ray band and hard lag in the hard X-ray band (their figure 11, the 1997 \#2 flare), in agreement with the fact that the X-ray spectrum in the soft band is flatter than in the hard band. This is consistent with the picture simulated by Kirk \et (1998).

If we interpret the lags as the difference of \tacc\ and \tcool\ at different energies (Zhang \et 2002)
\begin{equation}
\tau_{\rm hard} = t_{\rm acc}(E_{\rm h}) - t_{\rm acc}(E_{\rm l}) \,,
\label{eq:hardlag}
\end{equation}
\begin{equation}
\tau_{\rm soft} = t_{\rm cool}(E_{\rm l}) - t_{\rm cool}(E_{\rm h}) \,,
\label{eq:softlag}
\end{equation}
the observed lags are able to tell us the physical parameters of the emitting region (Zhang \et 2002)
\begin{equation}
B\delta \xi^{-2/3} = 0.21 \times (1+z) E_{\rm h}^{1/3} 
    \left [ \frac{1 - (E_{\rm l}/E_{\rm h})^{1/2}}
                 {\tau_{\rm hard}} \right ]^{2/3} \quad {\rm Gauss} \,,
\label{eq:hard}
\end{equation}
\begin{equation}
B\delta^{1/3} =209.91 \times \left (\frac{1+z}{E_{\rm l}}\right
)^{1/3}
	\left [\frac{1 - (E_{\rm l}/E_{\rm h})^{1/2}}
        {\tau_{\rm soft}} \right ]^{2/3} \quad {\rm Gauss}  \,.
\label{eq:soft}
\end{equation} 
where $\tau_{\rm hard}$ and $\tau_{\rm soft}$ refer to the observed hard and soft lags (in second) between the low $E_{\rm l}$ and high $E_{\rm h}$ energy (in keV), respectively ($E_{\rm l}$ and $E_{\rm h}$ are logarithmically averaged energies of the given energy bands). Equations~(\ref{eq:hard}) and~(\ref{eq:soft}) indicate that the larger the lag, the smaller the magnetic field $B$ of the emitting blob ($B \propto \tau^{-2/3}$, where $\tau$ means $\tau_{\rm soft}$ or $\tau_{\rm hard}$) if we assume that $\delta$ and $\xi$ of the system does not change significantly with time. As examples, we use \tcent\ derived from the CCFs of 0.2--0.8~keV versus 0.8--2.4~keV to estimate $B$ for the flares 362-1 and 450-3. We found that $B$ is $\sim 2.8\delta_{10}^{-1/3}$ and $0.6\delta_{10}^{-1/3}$ Gauss for 362-1 and 450-3, respectively, where $\delta_{10}$ means $\delta/10$. Therefore, the significant changes of soft lags suggest significant changes of $B$ with time. This phenomena was also observed in Mrk~421 (e.g., Zhang 2002; Ravasio \et 2004).

\begin{figure}
\plottwo{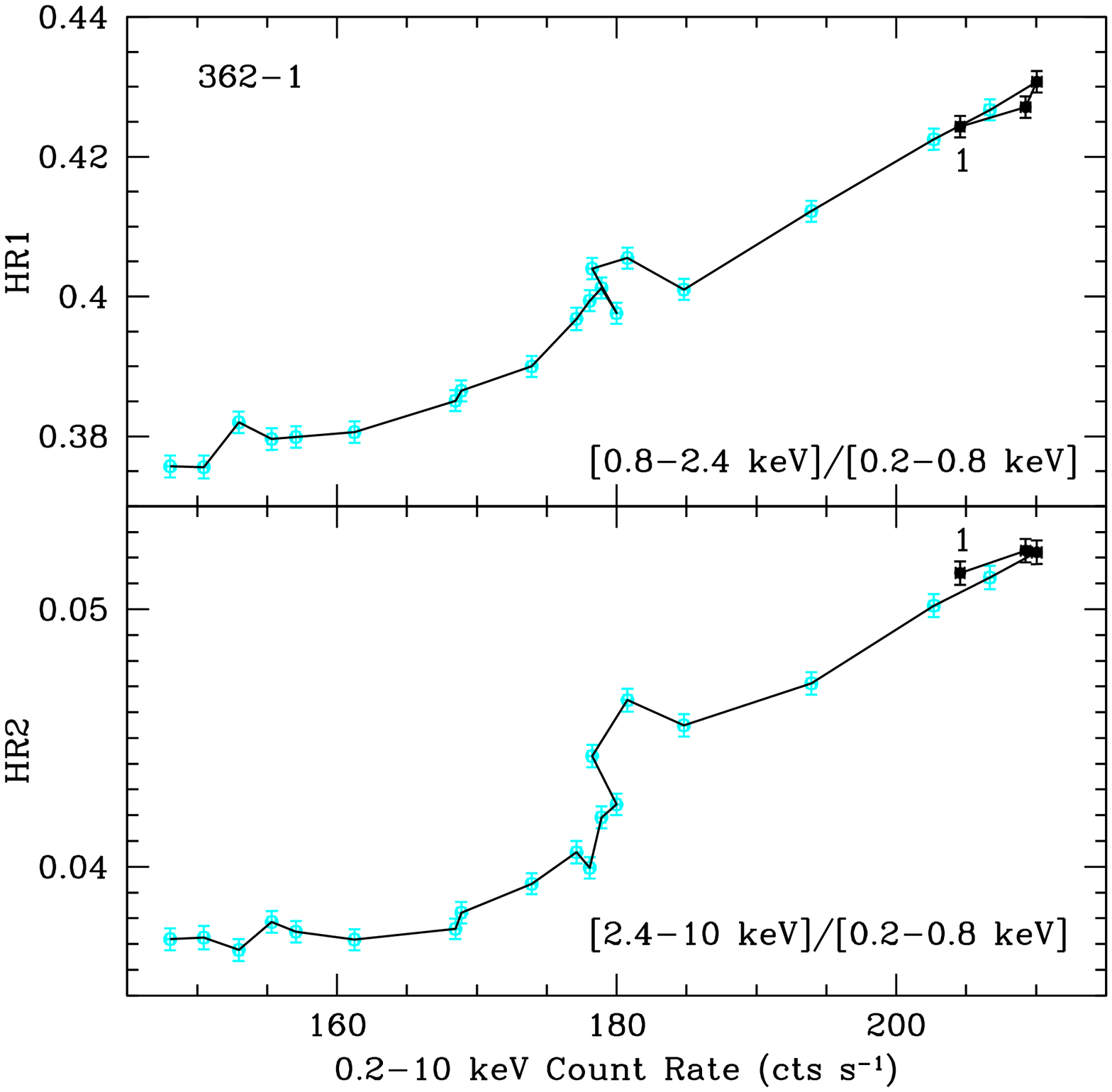}{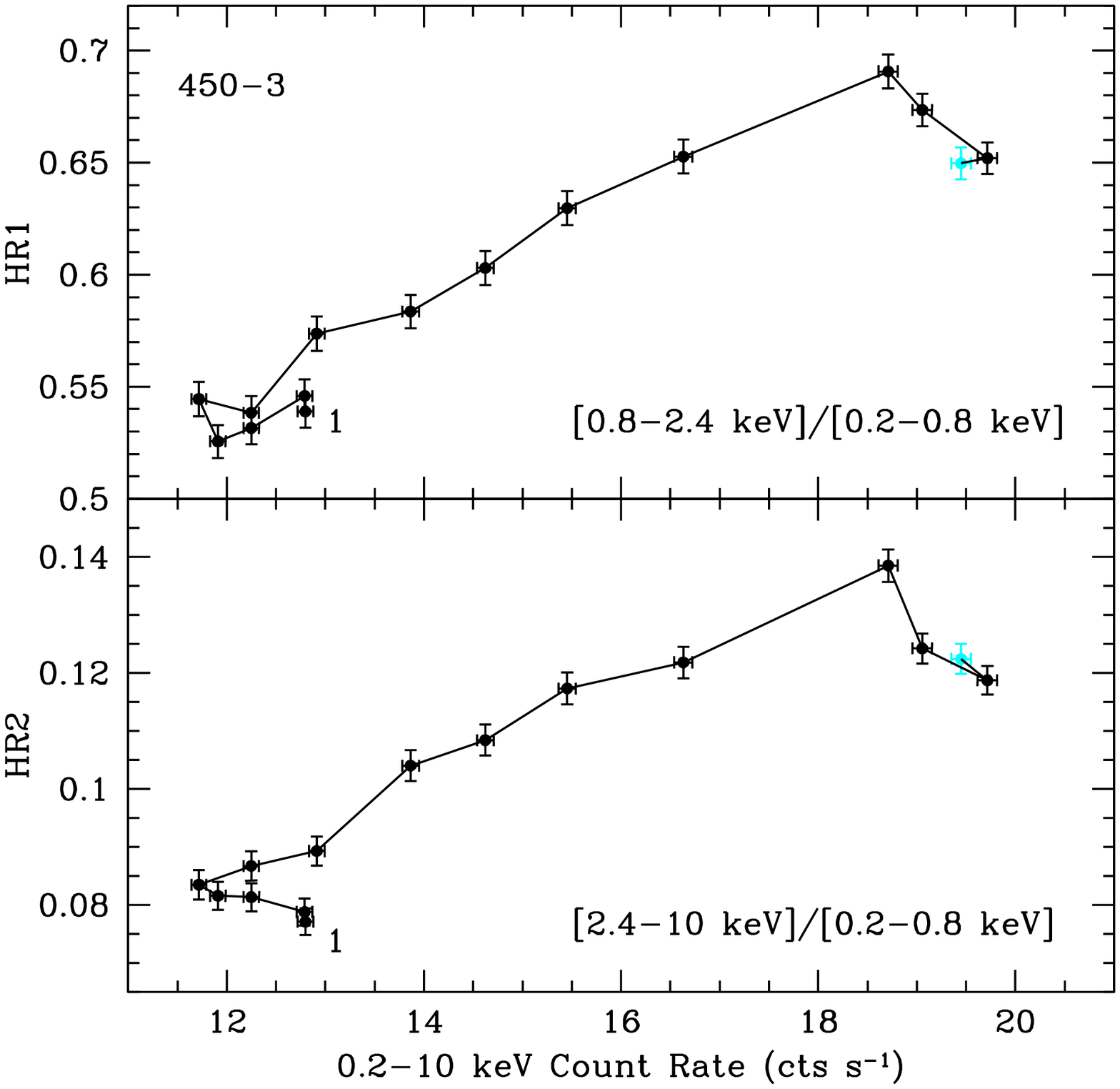}
\caption { \footnotesize The hardness ratios as a function of the 0.2--10~keV count rates. The data are binned in 2000~s. Dark points indicate the rising part of the flare, and the grey points the decaying part of the flare. The loops start from the point numbered with ``1'' and follow the connecting lines. The left plot is for the flare 362-1, and the right plot for the flare 450-3. The loops are not well-defined since both flares were not fully sampled. It appears that the flare 362-1 shows identical path while the flare 450-3 does not, for the rising and decaying phase of the flare, respectively. }
\label{fig:hrloop}
\end{figure}

The sign of a lag (either soft or hard lag) is determined by the balance between \tcool\ and \tacc\ of the electrons with same energy, while the amount of a lag is determined by the difference between the timescales (\tcool\ for soft lag and \tacc\ for hard lag) of the electrons with different energies. Equations~(\ref{eq:hardlag}) and~(\ref{eq:softlag}) show that the minor differences of the cooling or acceleration timescales between the emitting electrons with different energies may be not easily detected as lags. In case of soft lag, equation~(\ref{eq:soft}) indicates that the combination of the physical parameters $B\delta^{1/3}$ of the emitting region is large. Equations~(\ref{eq:tacc}) and~(\ref{eq:tcool}) further indicate that the acceleration/cooling timescales of the electrons are thus small. The changes of $\delta$ are not large, so $B$ must become large. For example, the lag (\tcent\ between the soft and medium bands) ratio of the flare 450-3 to 362-1 is about 10.3, from which we can infer that $B$ during the flare 362-1 is a factor of about 4.7 larger than that during the flare 450-3. The evolution of the flare 362-1 may thus be dominated by the light crossing time because of short \tcool, and the spectral evolution during the rising and decaying phase of the flare follows an identical path in the HR-count rate plot. In order to demonstrate this point, we re-binned the flares 362-1 and 450-3 data over 2000~s in the HR-count rate plots. The results are shown in Figure~\ref{fig:hrloop}. Because both flares were not fully sampled, it it difficult to compare the evolutionary paths of the rising and decaying phase of the flares. However, the left picture of Figure~\ref{fig:hrloop} appears to show that the paths of the rising and decaying phase of the flare 362-1 tend to be identical during the peak of the flare, consistent with the small lags derived with the CCFs. In contrast, the rising and decaying part around the peak of flare 450-3 do not follow an identical path and tend to be a clockwise loops on the HR-count rate plots, confirming large soft lags measured with the CCFs. Mrk~421 also showed similar behaviors (Brinkmann \et 2003). 


\section{Summary and Conclusions}\label{sec:conc}

With high signal-noise ratio and uniform light curves obtained with \xmm\ EPIC pn during 5 orbits over a period of about 3 years, we studied the spectral variability and measured the time lags of the bright TeV blazar \pks. The flux variability of the source was accompanied by significant spectral variability: the X-ray spectra hardens with increasing fluxes. We observed both soft and hard lags from seven exposures, and the delays between the 0.2--0.8~keV and 2.4--10~keV energy bands are larger than those between the 0.2--0.8~keV and 0.8--2.4~keV energy bands. Significant delays up to about one hour were also detected. The detections of time lags from this survey with \xmm\ observations are consistent with those obtained with \sax\ and \asca\ observations of \pks. The signs of the lags (either soft or hard lag) depend on the comparisons of particles cooling and acceleration timescales. The physical parameters of the emitting regions can be inferred from the amplitudes of lags.   


\acknowledgments
We thank the anonymous referee for the stimulating suggestions and comments that significantly improved the manuscript. This research is based on observations obtained with \xmm, an ESA science mission with instruments and contributions directly funded by ESA Member States and NASA. This work is conducted under Project 10473006 supported by National Natural Science Foundation of China, and partly under Project sponsored by the Scientific Research Foundation for the Returned Overseas Chinese Scholars, State Education Ministry.   


\clearpage
\input{tab1.tex}  
\clearpage
\input{tab2.tex}

\clearpage

\end{document}

%% file: tab1.tex
\begin{deluxetable}{rcccccccc}
\tablecolumns{9}
\tabletypesize{\footnotesize}
\tablewidth{0pt}
\tablecaption{The time lags (s)\tablenotemark{a}}
\tablehead{
\colhead{} &\multicolumn{4}{c}{Data} 
	 & &\multicolumn{3}{c}{Simulations\tablenotemark{b}}\\
\cline{2-5} \cline{7-9}
\colhead{Observations} 
	&\colhead{$r_{\rm max}$} 
	&\colhead{$\tau_{\rm peak}$} &\colhead{$\tau_{\rm cent}$}
	&\colhead{$\tau_{\rm fit}$}  &
	&\colhead{$\tau_{\rm peak}$} &\colhead{$\tau_{\rm cent}$}
        &\colhead{$\tau_{\rm fit}$}
}
\startdata
174-1 soft/medium  &0.94 &-600  &834  &-25 & 
     &$0^{+600}_{-900}$ &$562^{+600}_{-433}$ &$-18^{+270}_{-288}$ \\
      soft/hard    &0.87 &2100  &2860 &633 &
     &$1500^{+600}_{-1800}$ &$1947^{+1195}_{-889}$ &$572^{+683}_{-574}$ \\
362-1 soft/medium  &0.99 &0  &-156  &-470 & 
     &$0^{+0}_{-0}$ &$-156^{+7.6}_{-7.8}$ &$-466^{+119}_{-107}$ \\
      soft/hard    &0.97 &0  &-148 &-990 &
     &$0^{+0}_{-300}$ &$-151^{+11.8}_{-11.7}$ &$-910^{+261}_{-260}$ \\
362-2 soft/medium  &0.95 &300  &747  &549 & 
     &$300^{+600}_{-600}$ &$767^{+458}_{-456}$ &$393^{+382}_{-348}$ \\
      soft/hard    &0.76 &-300  &1605 &-230 &
     &$300^{+2100}_{-1200}$ &$1674^{+646}_{-763}$ &$...$\tablenotemark{c} \\
450-1 soft/medium  &0.86 &-900  &-457  &-157 &
     &$-300^{+600}_{-900}$ &$-467^{+313}_{-294}$ &$...$\tablenotemark{c} \\
      soft/hard    &0.61 &0  &-907 &-523 &
     &$-900^{+900}_{-1200}$ &$-872^{+606}_{-509}$ &$...$\tablenotemark{c} \\
%
%
450-3 soft/medium  &0.98 &-600  &-1744  &-815 &
     &$-900^{+600}_{-600}$ &$-1602^{+285}_{-295}$ &$-850^{+181}_{-180}$ \\
      soft/hard    &0.96 &-2100  &-3120 &-1353 &
     &$-1800^{+600}_{-900}$ &$-2824^{+424}_{-439}$ &$-1394^{+323}_{-359}$ \\
545-1 soft/medium  &0.93 &-2100  &-2381  &-1541 &
     &$-1800^{+600}_{-300}$ &$-2366^{+144}_{-149}$ &$-1402^{+191}_{-208}$ \\
      soft/hard    &0.78 &-3600  &-3443 &-3713 &
     &$-3600^{+1200}_{-1500}$ &$-3257^{+276}_{-187}$ &$-3154^{+834}_{-864}$ \\
724  soft/medium  &0.86 &-300  &-197  &-380 &
     &$-300^{+900}_{-1200}$ &$-307^{+458}_{-435}$ &$...$\tablenotemark{c}\\
\enddata
\tablenotetext{a}{The lags are not measured for the exposure 450-2 (no correlations) and the soft/hard CCF of the exposure 724 (no well-defined peak).}
\tablenotetext{b}{The quoted values are the medians of the simulations, and the
errors are $68\%$ confidence range with respect to the medians.}
\tablenotetext{c}{The fits are insensitive to the simulations.}
\label{tab:lag}
\end{deluxetable}

%% file: tab2.tex
\begin{deluxetable}{rccc}
\tablecolumns{9}
\tabletypesize{\footnotesize}
\tablewidth{0pt}
\tablecaption{Probability of detecting lags\tablenotemark{a}}
\tablehead{
\colhead{Observations} 
	&\colhead{$\tau_{\rm peak}$} 
	&\colhead{$\tau_{\rm cent}$} &\colhead{$\tau_{\rm fit}$}
	
}
\startdata
174-1 soft/medium  &42.3\%(S) &89.6\%(H) &52.6\%(S) \\
      soft/hard    &76.7\%(H) &98.5\%(H) &84.0\%(H) \\
362-1 soft/medium  &100\%(0) &100\%(S) &100\%(S) \\
      soft/hard    &96.2\%(0) &100\%(S) &99.9\%(S) \\
362-2 soft/medium  &59.8\%(H) &96.0\%(H) &87.3\%(H) \\
      soft/hard    &50.4\%(H) &98.3\%(H) &... \\
450-1 soft/medium  &54.9\%(S) &92.4\%(S) &... \\
      soft/hard    &67.8\%(S) &91.7\%(S) &... \\
450-3 soft/medium  &93.8\%(S) &100\%(S) &100\%(S) \\
      soft/hard    &98.3\%(S) &100\%(S) &100\%(S) \\
545-1 soft/medium  &100\%(S) &100\%(S) &100\%(S) \\
      soft/hard    &100\%(S) &100\%(S) &100\%(S) \\
724  soft/medium  &70.0\%(S) &74.7\%(S) &... \\
\enddata
\tablenotetext{a}{The maximum probability,  calculated with the simulation values, that the lag is either soft (S), hard (H) or zero (0) lag. }
\label{tab:prob}
\end{deluxetable}